\documentclass[letterpaper,11pt]{article}

\usepackage{jheppub} 

\usepackage{slashed}
\usepackage{graphicx}
\usepackage{subfigure}

\newcommand{\beq}{\begin{equation}}
\newcommand{\eeq}{\end{equation}}
\newcommand{\bea}{\begin{eqnarray}}
\newcommand{\eea}{\end{eqnarray}}

\title{Towards the fate of natural composite Higgs model through single $t^\prime$ search at the 8 TeV LHC}

\author[a]{Jinmian Li,}
\author[a]{Da Liu,}
\author[a]{Jing Shu}

\affiliation[a]{State Key Laboratory of Theoretical Physics, Institute of
Theoretical Physics, Chinese Academy of Sciences,Beijing 100190, People's Republic of China.\footnote{jshu@itp.ac.cn}}

\abstract{We analyze the observational potential of single $t'$ production in both the $t^\prime \rightarrow bW$ and $t^\prime \rightarrow th$ decay channels at 8 TeV LHC using an integrated  luminosity of 25 $\text{fb}^{-1}$. Our analysis is based on a simplified model with minimal coset $SO(5)/SO(4)$ in which the $t'$ is a singlet of the unbroken $SO(4)$. The single $t'$ production, as a consequence of electroweak symmetry breaking, is less kinematically suppressed, associated with a light forward jet and has boosted decay products at the 8 TeV LHC. Therefore it provides the most promising channel in searching for a heavy $t'$. We have exploited the above kinematical features and used the jet substructure method to reconstruct the boosted Higgs in $th$ decay channel. It is shown that a strong constraint on the $t^\prime bW$ coupling ($g_{t^\prime bW}/g_{tbW,SM} < 0.2 \sim 0.3$) at the 95\% C. L. can be obtained for $m_{t'} \subset (700, 1000)$ GeV. 
}

\begin{document}

\maketitle


\section{Introduction}

The discovery of a 125 GeV Higgs-like boson last year at the Large Hadron Collider motivates us to reconsider the origin of electroweak symmetry breaking (EWSB) in the next few decades. One long related outstanding question is the naturalness problem: why we have such a dramatic cancellation on the radiative Higgs effective potential which sensitively depends on the ultraviolet physics. Popular models have been proposed and different solutions can be categorized simply based on the objects which cancel the largest Higgs radiative corrections from the Standard Model (SM) top quark. Light scalar particles (stops) would be essential for natural supersymmetry scenario while vector-like quarks with their contributions from higher dimensional operators would appear in composite Higgs models (CHMs).

The second solution, the existence of vector-like quarks, is related to another historical question that why should we only have three generations of fermions. Another chiral quark which is an exact copy of light generation, is highly constrained by indirect searches like electroweak precision test~\cite{Erler:2010sk, Eberhardt:2010bm, Murayama:2010xb}, Higgs production and decay~\cite{Kribs:2007nz, Keung:2011zc, Eberhardt:2012ck, Eberhardt:2012gv, Carpenter:2011wb}. Vector-like quarks \footnote{It is interesting to notice that extra vector-like quarks can also help to explain the Higgs mass and the muon $g-2$ anomaly in the context of supersymmetry \cite{Endo:2011xq}.}, nevertheless, are very weakly constrained from the above measurements because their contributions are suppressed by the large vector masses. 

More recently, an interesting observation has been made in the context of CHMs on the vector-like quark masses~\cite{Matsedonskyi:2012ym, Berger:2012ec, Redi:2012ha, Marzocca:2012zn, Pomarol:2012qf, Panico:2012uw, Pappadopulo:2013vca}. If the EWSB is triggered by the radiative Higgs potential from the top quark which mixes linearly with the composite operator that consists of vector-like quarks, there is a upper bound for the lightest vector-like quark for a 125 GeV composite Higgs for fixed $\xi$ (roughly $<$ 900 GeV for $\xi=0.1$). This provide a very complimentary approach to test CHMs besides the Higgs coupling deviations which depends on $\xi = v^2 / f^2$ at the leading order. Therefore, the genuine smoking gun of CHMs is not only the deviation of composite Higgs couplings from SM ones, but also the existence of light vector-like quarks \footnote{We notice that by considering the $SU(5)/SO(5)$ breaking pattern \cite{Vecchi:2013bja}, the deviation of the composite Higgs couplings to the SM ones could be much smaller than the case of minimal CHM. Therefore, this provide more support to test CHMs in the top partner searches. }.



If the lightest vector-like quark is a mixture of top quark and a $SO(4)$ singlet, then its charge is $2/3$ and serves as a $t'$ particle. Unlike a chiral $t'$ which primarily decays into $b W^+$, the vector $t'$ has large flavor changing neutral couplings to Higgs and $Z$ boson and its decay branching ratio into $b W^+$, $th$ and $tZ$ is $2:1:1$ in the Goldstone equivalence limit ($m_{t'} \rightarrow \infty$). Therefore, a comprehensive study combing all the decay channels or at least two would be helpful to get either a strong constraint or a hint for discovery. 

The pair production of top partners at the LHC has been searched at CMS which sets the $t'$ mass bound of 570 GeV or 625 GeV at 95\% confidence level (C. L.) assuming purely $bW$ or $tZ$ decay~\cite{Chatrchyan:2012vu,Chatrchyan:2012af}, while at ATLAS, the bound is 656 GeV at 95\% CL for pure $bW$ decay channel~\cite{ATLAS:2012qe}. The discovery of Higgs-like boson has made the $th$ decay channel promising and it has been considered in the multi-b jet final state~\cite{Harigaya:2012ir, Girdhar:2012vn, Vignaroli:2012nf}. A lower bound of 640 GeV was recently set on the $t'$ mass \cite{ATLAS:Tth} at ATLAS using high multiplicity of $b$ jets of this channel with at least one Higgs boson decaying into $b \bar{b}$. Other channels using the muti-b jets final states in stop decay $\tilde{t} \rightarrow t h \tilde{\chi}$ \cite{Berenstein:2012fc} and $t' \rightarrow h^+ b$ \cite{Kearney:2013oia} have also been investigated.

We notice however, for 8 TeV LHC, search for heavy $t'$ ($>$ 600 GeV) from the single production via electroweak couplings could be more promising than the pair production via QCD.
First, the single production is less kinematically suppressed by the large $t'$ mass. Second, the extra jet from splitting of a valence quark with one $W$ emission always has a strong forward nature. Third, the next decay products from heavy $t'$ has a large space separation where the further decay products are highly collimated. With all those features, we expect to have a large signal to backgrounds ratio in the $t'$ single production channel. Therefore, in this paper we study the observability of a single $t^\prime$ production at 8 TeV LHC combing both the $bW$ and $th$ decay channels. The $tZ$ channel has either small number of events in the $Z$ di-lepton decay channel or large backgrounds in the $Z$ hadronic decay channel. For $th$ decay channel, jet substructure method is  applied to reconstruct the hadronically decaying Higgs boson because of  its moderately boost  in the relatively high mass region of the $t^\prime$.

The rest of the paper is organized as follows. In Seciton~\ref{sec:setup} we describe the main features of  the simplified model, in a similar fashion as~\cite{DeSimone:2012fs}.  In Section~\ref{sec:singlechannel} we discuss the single $t^\prime$ production at the 8 TeV LHC and its decay channels that we consider in this paper. In Section~\ref{sec:Event} we turn to study the prospects of observing the single $t^\prime$ production by performing a detailed analysis of   the signal and backgrounds in each channel. We conclude and give a outlook in Section~\ref{sec:conclusion}.

\section{Simplified model based on $SO(5)/SO(4)$}
\label{sec:setup}
Following \cite{DeSimone:2012fs}, we consider a simplified composite Higgs model based on $SO(5)/SO(4)$, where the right-handed top quark $t_R$ belongs to the singlet of $SO(4)$ in the strongly interacting sector. Top Yukawa is generated by the linearly coupling between $q_L = (t_L, b_L)^T$ and the composite operators, according to the partial compositeness scenario \cite{Kaplan:1991dc}. The composite top  partners can be either in the fourplet or singlet of the unbroken $SO(4)$. While the lightest top partner from the fourplet is the exotic charge $5/3$, we focus on the singlet case as the possible lightest top partner with charge 2/3, $t^\prime$. The SM elementary fields are embedded as fundamental representation of $SO(5)$ $\xi_L$, which formally transforms as $\xi_L\rightarrow g \xi_L$ under $g\in SO(5)$ in the spurion language:
  \beq
  \Psi = \left(\begin{array}{c}
              0\\
              0\\
              0\\
              0\\
              t^\prime
              \end{array} \right)_{\frac23},\qquad
\xi_{L}=\frac{1}{\sqrt{2}}
\left(\begin{array}{c}
ib_{L}\\
b_{L}\\
it_{L}\\
-t_{L}\\
0
\end{array}\right)_{\frac{2}{3}},\qquad
\xi_{R}=\left(\begin{array}{c}
0\\
0\\
0\\
0\\
t_{R}\end{array}\right)_{\frac{2}{3}},
  \eeq
where $2/3 $ is $U(1)_X$ charge in order to reproduce the right electric charge of the SM fields $Q = T^{3L} +  T^{3R} + X$, $T^{3(L,R)}$ are the third $SO(4)$ $\simeq$ $\text{SU}(2)_L \times \text{SU}(2)_R$ unbroken generators. The above way of embedding is not the unique way but can simply protect the tree level $Z \bar{b}_L b_L $ vertex \cite{Agashe:2006at}. 

Based on Callan- Coleman-Wess-Zumino (CCWZ) construction \cite{Coleman:1969sm}, the general effective Lagrangian formally invariant under $SO(5) \times U(1)_X$ to the leading order is:
 \bea
 \mathcal{L}&=&\bar{q}_Li\slashed D q_L + \bar{t}_Ri\slashed D t_R + \bar{\Psi}i\slashed D \Psi-M_\Psi \bar{\Psi}\Psi\nonumber\\
            &+&\epsilon_{qt^\prime}\bar{\xi}_LU \Psi_R + \epsilon_{qt}\bar{\xi}_LU \xi_R + h.c. .
  \label{Lag}
 \eea
where $U$ is the Goldstone boson $5\times5$ matrix,
\beq
U= \text{exp}(i\frac{\sqrt{2}}{f}h^{\hat{a}}T^{\hat{a}})
=\left(
  \begin{array}{cc}
    \textbf{1}_{4\times4}-\frac{\vec{h} \vec{h}^T}{h^2}(1-\cos\frac{h}{f}) & \frac{\vec{h} }{h}\sin\frac{h}{f} \\
    -\frac{\vec{h}^T}{h}\sin\frac{h}{f} & \cos\frac{h}{f} \\
  \end{array}
\right)
\label{U}
\eeq
and takes a simple form in unitary gauge:
\beq
U=\left(\begin{array}{ccccc}
         1  &  0 & 0  & 0  & 0\\
         0  &  1 & 0  & 0  & 0\\
         0  &  0 & 1  & 0  & 0\\
         0  &  0 & 0  & \text{cos}\frac{h}{f}  & \text{sin}\frac{h}{f}\\
         0  &  0 & 0  & -\text{sin}\frac{h}{f} & \text{cos}\frac{h}{f}\\
        \end{array} 
\right) ,
\label{Uuni}
\eeq
where we have used the same notation h for 
$\sqrt{h^{\hat{a}}h^{\hat{a}}}$ in eq.~(\ref{U}) and for the physical Higgs field in eq.~(\ref{Uuni}).
Note that the  elementary-composite interactions of eq.~(\ref{Lag}) break the $SO(5)$ explicitly and will contribute to the Higgs potential. We have neglected the direct mixing term among $t_R$ and $t^\prime_R$ which can be removed by a field redefinition. There are three parameters ($\epsilon_{qt^\prime},\epsilon_{qt}, M_\Psi$) in addition to the decay constant $f$ in Goldstone matrix and all of them can be made real by phase rotation of the chiral fields. After EWSB, the first term in the second line will induce a mass mixing between the top quark and the $t'$ and the mass matrix in the basis ($t,t^\prime$) is simple:
 \beq
 \left(\begin{array}{cc}
       \bar{t}_L  &\bar{t^\prime}_L
       \end{array}\right)
 \left(\begin{array}{cc}
 \frac{\epsilon_{qt}}{\sqrt{2}} \sqrt{\xi} & \frac{\epsilon_{qt^\prime}}{\sqrt{2}} \sqrt{\xi}\\
 0                                         & M_\Psi
 \end{array}\right)
 \left(\begin{array}{c}
        t_R\\
        t^\prime_R
       \end{array}
\right) ,
 \eeq
where $\xi= \text{sin}^2\frac{<h>}{f}=(\frac{v}{f})^2$ which is smaller than 0.2 as suggested by Electroweak precision test. Note there is a zero element in the matrix since the right-handed top quark $t_R$ is a composite singlet in our case. This remarkable feature of the mass matrix will simplify our calculation and we can diagonalize it simply by chiral rotation of the top quark and the top partner fields:
\beq
\left(\begin{array}{c}
 t_L\\
 t^\prime_L
\end{array}\right)
\rightarrow
\left(\begin{array}{cc}
 \text{cos}\alpha & \text{sin}\alpha\\
 -\text{sin}\alpha & \text{cos}\alpha
\end{array}
\right)
\left(\begin{array}{c}
 t_L\\
 t^\prime_L
\end{array}\right),\qquad
\left(\begin{array}{c}
 t_R\\
 t^\prime_R
\end{array}\right)
\rightarrow
\left(\begin{array}{cc}
 \text{cos}\beta & \text{sin}\beta\\
 -\text{sin}\beta & \text{cos}\beta
\end{array}\right)
\left(\begin{array}{c}
 t_R\\
 t^\prime_R
\end{array}\right) \ .
\eeq

It is convenient to rewrite the Lagrangian parameters in terms of physical parameters ($\alpha, \beta,m_t, m_{t^\prime}$) and use them later in our analysis. The zero element in the mass matrix allows us to further rewrite the sine of right-handed mixing angle $\beta$ in terms of a function of ($\alpha, m_t, m_{t^{\prime}}$):
      \bea
     \epsilon_{qt}& = & \frac {\sqrt{2}}{\sqrt{\xi} }\frac{m_t m_{t^\prime}}{\sqrt{m_t^2\text{sin}^2\alpha+m_{t^\prime}^2\text{cos}^2\alpha}},\nonumber\\
     \epsilon_{qt^\prime}& =& \frac {\sqrt{2}}{\sqrt{\xi} }\frac{(m_{t^\prime}^2-m_t^2)\text{sin}\alpha\text{cos}\alpha}{\sqrt{m_t^2\text{sin}^2\alpha+m_{t^\prime}^2\text{cos}^2\alpha}},\nonumber\\
     M_\Psi &=&\sqrt{m_t^2\text{sin}^2\alpha+m_{t^\prime}^2\text{cos}^2\alpha},\nonumber\\
      \text{sin}\beta&=&\frac{m_t\text{sin}\alpha}{\sqrt{m_t^2\text{sin}^2\alpha+m_{t^\prime}^2\text{cos}^2\alpha}}
     \eea

\begin{figure}[ht]
$\includegraphics[width=0.5\textwidth]{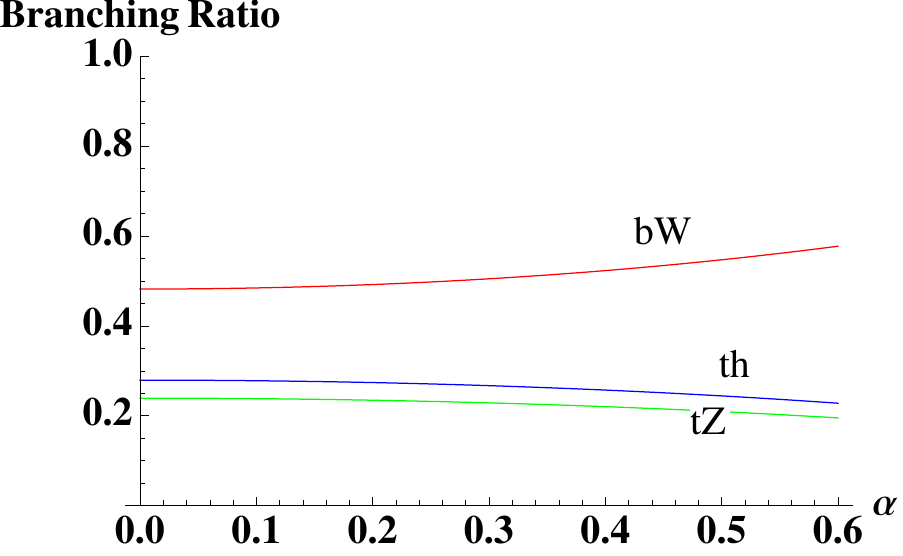}$ 
$\includegraphics[width=0.5\textwidth]{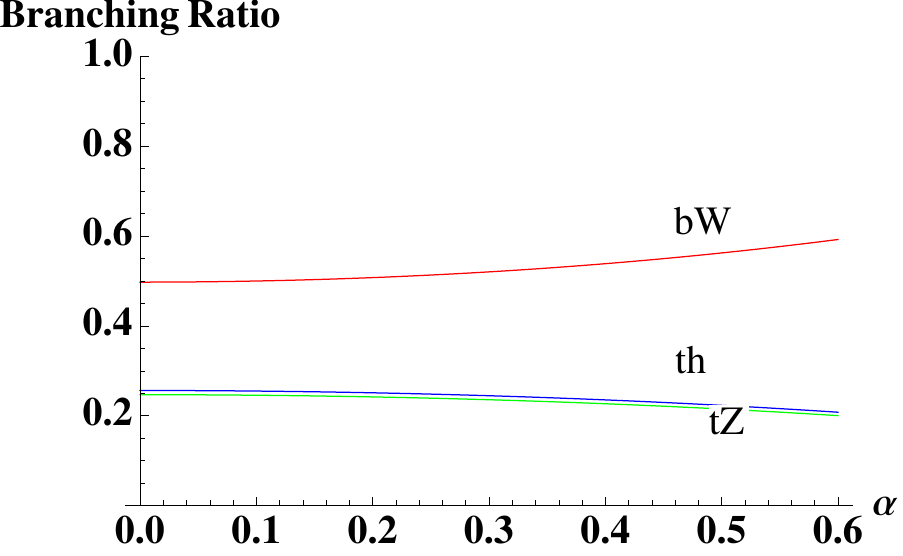}$ 
\caption{Branching ratios for the three decay modes of $t'$ with $m_{t^\prime}$ = 700 GeV. Left panel: $\xi =0.1$. Red line: $bW$ decay channel; blue line: $th$ decay channel; green line: $tZ$ decay channel. Right panel:  The same plot with $\xi$ = 0.2.}
\label{BR}
\end{figure}

Since we consider the relatively high mass region of the top partner, it is convenient to study its dynamics by using the Goldstone boson equivalence theorem. To get the couplings among the top partner and the charged, neutral Goldstone fields $\phi^{\pm}, \phi^0$, we first rewrite the fourplet $\vec{h}$ in terms of the standard Higgs doublet of $+\frac12$ hypercharge:
\beq
\vec{h}=\left(\begin{array}{c}
  h_1 \\
  h_2 \\
  h_3 \\
  h_4 
\end{array}\right)
= \left(\begin{array}{c}
             \frac{-i}{\sqrt{2}}(\phi^+-\phi^-) \\
             \frac{1}{\sqrt{2}}(\phi^++\phi^-) \\
             -\phi^0 \\
                h
           \end{array}\right). 
\eeq
Then it is straightforward to obtain the Goldstone couplings by using the explicit form of $U$ matrix in eq. (\ref{U}). Neglecting
the EWSB, the mixing term between $q_L$ and $t^\prime$ in the effective Lagrangian gives:

 \beq
 \epsilon_{qt^\prime}\bar{\xi}_LU \Psi_R \sim  -\frac{\epsilon_{qt^\prime}}{\sqrt{2}f}(h-i\phi^0)\bar{t}_Lt^\prime_R
 + \frac{\epsilon_{qt^\prime}}{f} \phi^{-}\bar{b}_L t^\prime_R
 \eeq
 Note there is a $\sqrt{2}$ suppression of the coupling with the top quark, from which
  we can easily obtain  BR($t^\prime\rightarrow t h$) $\approx$ BR($t^\prime\rightarrow t Z$) $\approx$ 
 BR($t^\prime\rightarrow b W$)/2 $\approx$ 0.25 in the gauge eigenstate. One can also calculate the 
partial decay widths explicitly and obtain the exact formulas \cite{Perelstein:2003wd} : 
\bea
\Gamma_{bW}&=&\frac{g^2\text{sin}^2\alpha m_{t^\prime}^3}{64 \pi m_W^2} f(x_W,x_b)
g(x_b,x_W),\nonumber\\
\Gamma_{tZ}&=&\frac{g^2\text{sin}^2\alpha \text{cos}^2\alpha m_{t^\prime}^3}{128 \pi m_W^2} f(x_Z,x_b)
g(x_b,x_Z),\nonumber\\
\Gamma_{th}&=&\frac{ g^2 \text{sin}^2\alpha \text{cos}^2\alpha m_{t^\prime}^3 (1-\xi)}{128 \pi m_W^2} f(x_t,x_h)\left[ (1 + x_t^2
 - x_h^2)(1+x_t^2) + 4x_t^2\right],
\eea
where $x_i$ are defined as $x_i = \frac{m_i}{m_{t^\prime}}$, the kinematic functions 
are given by:
\bea
f(x_i,y_j) &=& \sqrt{(1-(x_i + x_j)^2)(1 - (x_i - x_j)^2)},\nonumber\\
g(x_i,y_j) &=& 1 - x_i^2 + x_j^2(1 + x_i^2) - 2x_j^4.
\eea
In the large mass limit of the top partner $m_{t^\prime} \rightarrow \infty$, $x_i \rightarrow 0$, without the mixing effects, one can see that $\Gamma_{bW} : \Gamma_{tZ} : \Gamma_{th}$ = 2 : 1 : 1 from the above expressions in the limit $\xi \rightarrow 0$.

\begin{figure}[ht]
$\includegraphics[width=0.8\textwidth]{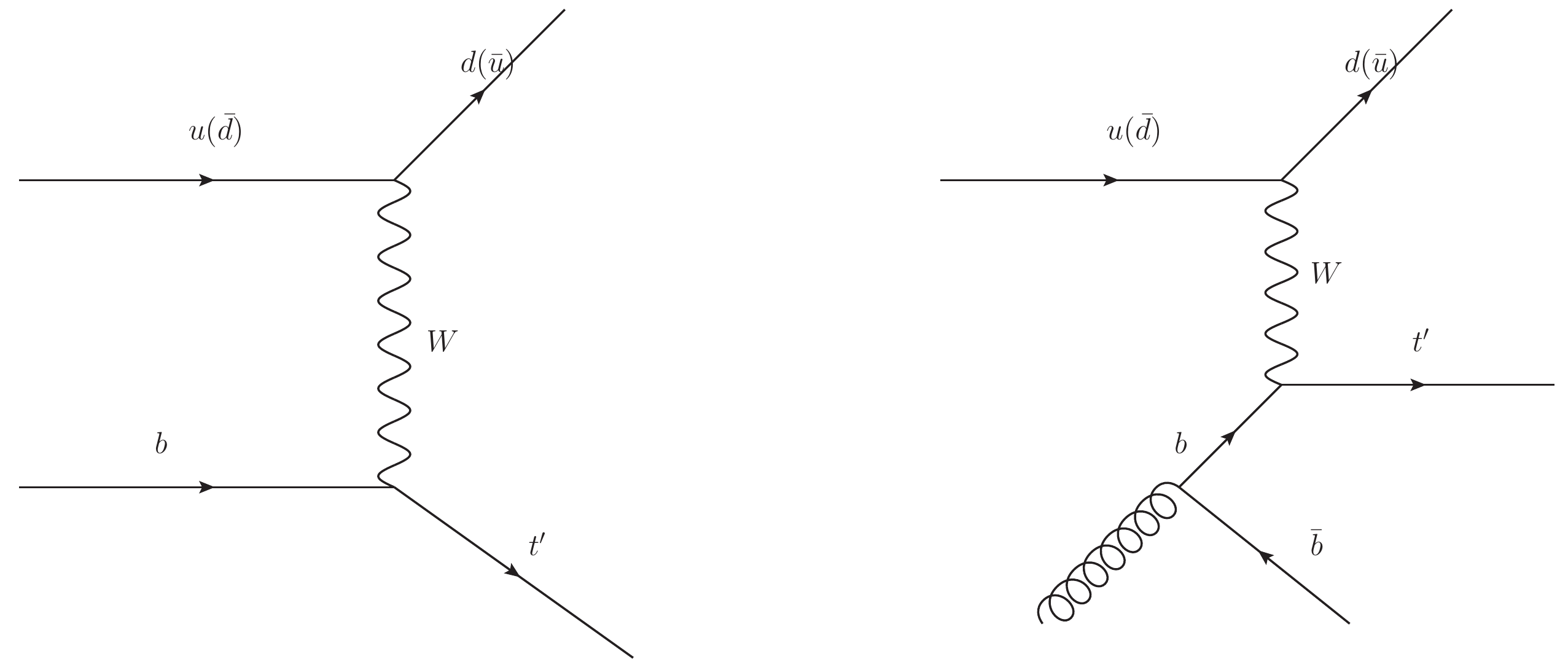}$
\caption{Feynman diagrams of the t-chanel single $t^\prime$ production. The right diagram is the NLO correction in our calculation which will produce an additional b-quark. See text for detail.}
\label{singletp}
\end{figure}
  
We show the branching ratios of three decay channels by varying left-hand mixing angle  $\alpha$ with $m_{t^\prime}$ = 700 GeV in Figure \ref{BR}, from which we can see either $\xi=0.1$ or $\xi=0.2$, BR($t^\prime\rightarrow t h$) $\approx$ BR($t^\prime\rightarrow t Z$) $\approx$ 
$\frac12$BR($t^\prime\rightarrow b W$) $\approx$ 0.25 is a good approximation as expected by Goldstone boson equivalence theorem.
We will use this branching fractions in our analysis and $m_{t^{\prime}}$ = 700 GeV as our benchmark point.

\section{The single $t^\prime$ production}
\label{sec:singlechannel}

The single $t^\prime$ can be produced via its electroweak interactions and the leading process is the t-channel which originates from the $W-b$ fusion where $W$ is emitted by a light quark in the proton \footnote{There is another possibility that the new vector-like quarks mixe sizably with the SM light quarks (See~\cite{Atre:2008iu}), but then their masses are not connected to EWSB. We do not consider this case in our paper although their production cross section will be very large due to the mixing with valence quarks. }. The cross section of single 
$t^\prime$ production has been calculated recently up to NLO using MCFM code 
\cite{Campbell:2009gj} in a scheme with the proton containing four flavors of quark where there will be an additional bottom quark produced from
the gluon splitting~\cite{DeSimone:2012fs}. Given that this bottom quark is very soft at 8 TeV LHC, we just omit it and recalculate the cross
section using b-quark parton distribution function in the proton. The corresponding Feynman diagrams are shown in Figure
\ref{singletp} and the gluon-splitting one will be the NLO correction of the single production. 

\begin{figure}[ht]
\centering
$\includegraphics[width=0.6\textwidth]{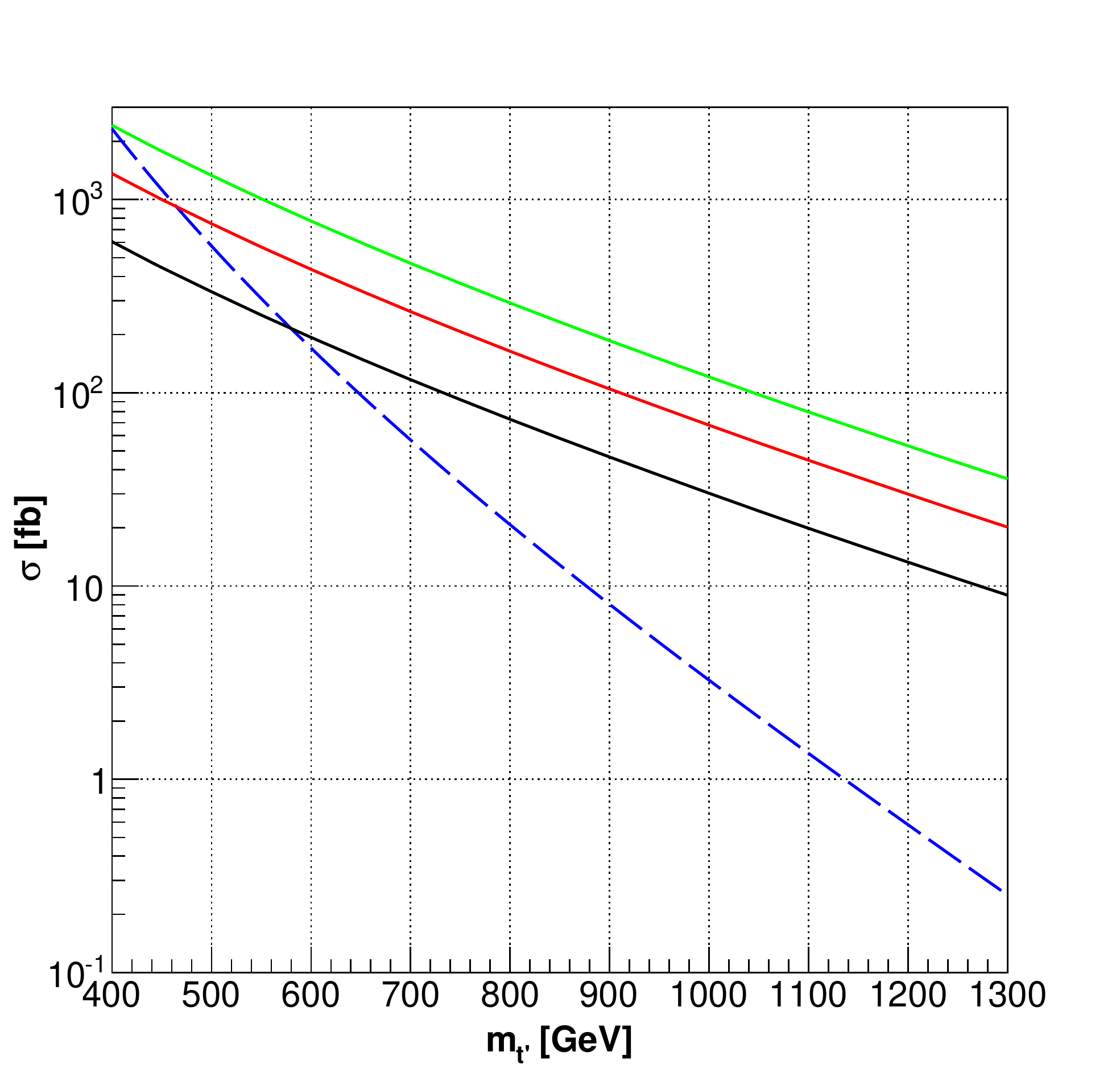}$
\caption{Cross sections of electroweak single $t^\prime$ productions and QCD pair production. The solid black, red, green or dashed blue line stands for the electroweak single production with sin$\alpha$ = 0.2, 0.3, 0.4, or QCD pair production, respectively. }
\label{XS}
\end{figure}

We plot in Figure \ref{XS} the cross sections for QCD pair production and electroweak single production of $t^\prime$ with $\sin \alpha$ = 0.2, 0.3, 0.4 as a function
of its mass . For the $t^\prime$ pair production, we use the HATHOR code \cite{Aliev:2010zk}
which includes perturbative QCD corrections up to NNLO. In all the calculations, we choose the set of the parton distribution
functions MSTW2008 \cite{Martin:2009iq}. We see that single production rate dominates over pair production in the moderately 
high mass region with appropriate $t^\prime bW$ coupling which is just $\frac{g}{\sqrt{2}}$sin$\alpha$ in the singlet case.
This is due to the lower kinematical threshold of single $t^\prime$ production in comparison with the pair production.

Another advantage of considering single $t^\prime$ production channel is that since the cross section is proportional to the square of $t^\prime bW$ coupling and the branching ratios for the three different decay channels are almost fixed at the relatively high mass as
discussed in sect.~\ref{sec:setup}, these search channels can be treated as a direct probe of the $t' b W$ coupling. 

Concerning the $t^\prime$ decay, we consider the $bW$ and $th$ decay channels in this paper and study their practical observability by analysing the signal and backgrounds respectively. For the $th$ decay channel, we take the advantage of multi-$b$ signature of the signal from Higgs decay and its large transverse momentum magnitude so that jet substructure method can be used to tag the boosted Higgs. Additionally, leptonic decay of the top will be  considered in order to suppress the overwhelming QCD multi-jet backgrounds. We do not consider the $tZ$ decay channel here because of the small number of events for leptonic $Z$ decay, although one can use the top tagging method to improve the discovery potential. 



\section{Event generation and  analysis}
\label{sec:Event}
In this section, we analyze the observation potential of each channel by performing a Monte Carlo
simulation of the signal and background events and applying the suitable selection cuts to 
maximize the significance. We will take $m_{t^\prime} = $700 GeV as our benchmark scenario to 
illustrate event reconstruction techniques used in the analysis. Results of different masses of the top partner are shown separately assuming sin$\alpha$ = 0.2 and sin$\alpha$ = 0.3 for $bW$ decay channel, while for $th$ decay channel the value of sin$\alpha$ = 0.4 is adopted.  We finally
derive the expected 95\% C. L. exclusion plots in the $m_{t^\prime}-\text{sin}\alpha$ plane for each channel and for combination of the two channels. 

\subsection{$bW$ decay channel}
\label{bW}
In this section, we consider the $bW$ decay channel of the top partner $t^\prime$ with subsequent  
decay of $W$ boson into $\ell\nu$, where $\ell$ is either an electron or a muon. The largest background comes from  the $W$ + light jets with one of the jets misidentified as a $b$-quark jet. The cross section is normalized as $\sim$ 3.96 nb from MadGraph5~\cite{Alwall:2011uj} LO calculation multiplied with a K-factor of  1.2, where the jet partons are required to have $p_T > $ 30 GeV. $W + b+$ light jets and $W$ + $b\bar{b}$ can also make contributions to the backgrounds because they contain real $b$-quarks. The cross sections are 38.4 pb and 11.5 pb obtained similarly with $W$ + light jets. The second important background is $t\bar{t}$ with cross section $\sim$ 238 pb as obtained from  approximate NNLO QCD calculations with HATHOR~\cite{Aliev:2010zk}. 
Other smaller backgrouds come from single top \cite{Kidonakis:2011wy,Kidonakis:2010ux,Kidonakis:2010tc}  and diboson ($WW$, $WZ$) production.

The interactions of $t'$ are implemented into an UFO file for MadGraph5~\cite{Alwall:2011uj} by using  the Feynrules~\cite{Feynrule} package based on the simplified model. Parton-level events for the signal and backgrounds are generated by MadGraph5~\cite{Alwall:2011uj} and interfaced to PYTHIA-6.420~\cite{Sjostrand:2006za} for 
parton-showering and hadronization, after which Delphes 3.0.5 \cite{Ovyn:2009tx} is used for detector simulation. The decay of SM particles (top quark and $W$ boson) are performed in PYTHIA 6.420. Jets are reconstructed using anti-$k_t$ algorithm \cite{Cacciari:2008gp} with a radius parameter $R = 0.4$ and required to satisfy $|\eta|< 4.5 $ and $p_T > 30$ GeV. For the  b-quark jets, only the ones satisfying $|\eta|< 2.5$ are considered. In addition, we take
the b-tagging efficiency of 70\% for b-quark jets, miss-b-tagging probability of 1\% for light jets and 20\%  for c-quark jets \cite{ATLAS:Tth}. Furthermore, leptons are required to have $|\eta|< 2.5 $, $p_T > 25$ GeV and be isolated. By``isolated", we mean
the scalar $p_T$ sum of the tracks and the calorimeter cells within a cone of $R=0.2$ around the lepton is no more than 10\% of the lepton transverse momentum, where the calculation is done by  simulating the energy flow algorithm. 

\begin{figure}[ht]
 \centering
 \subfigure[]{
    \label{bW:a}
    \includegraphics[width=0.45\textwidth]{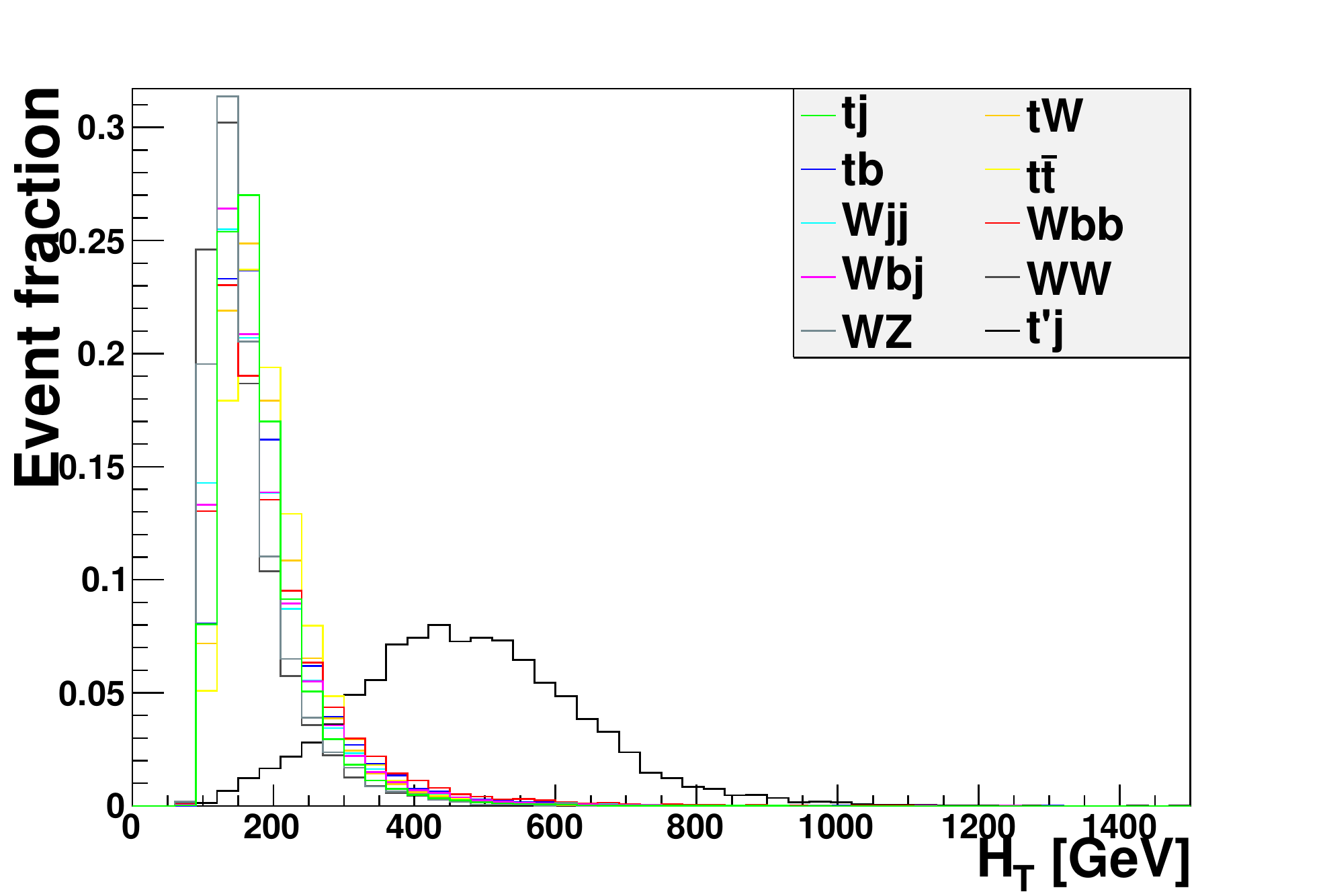}}
\hspace{0.05in}
 \subfigure[]{
    \label{bW:b}
 \includegraphics[width=0.45\textwidth]{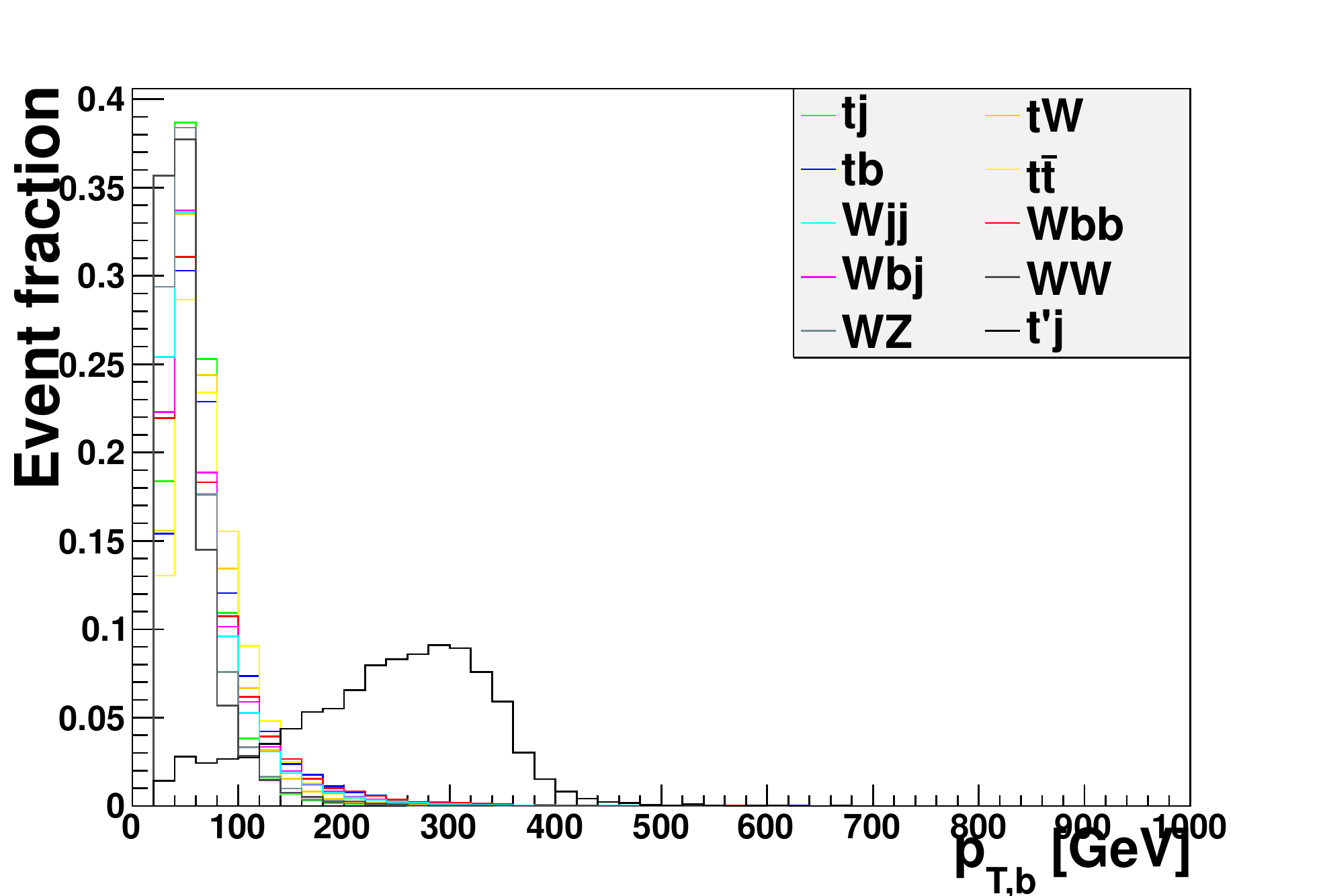}}

 \subfigure[]{
    \label{bW:a}
    \includegraphics[width=0.45\textwidth]{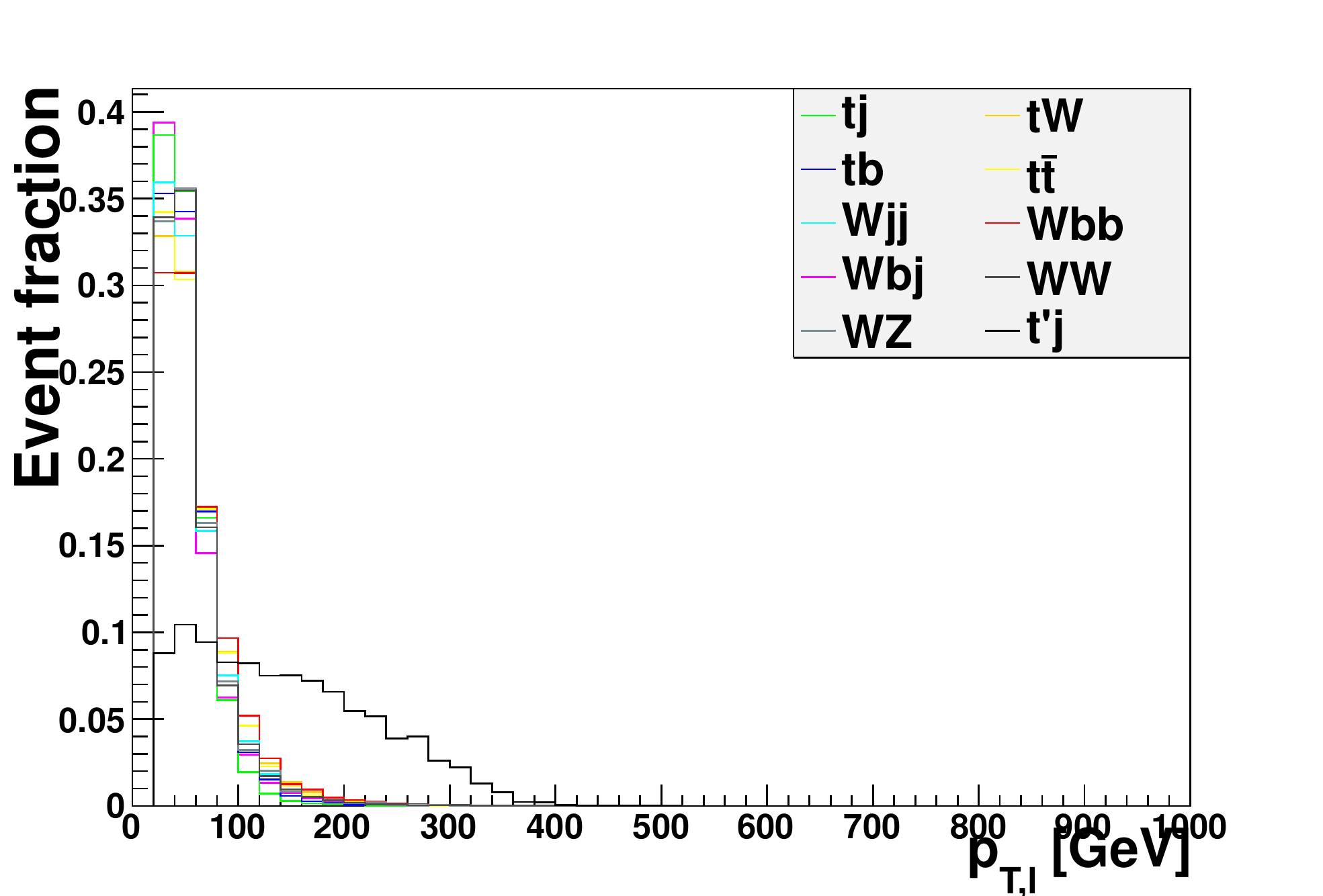}}
\hspace{0.05in}
 \subfigure[]{
    \label{bW:b}
 \includegraphics[width=0.45\textwidth]{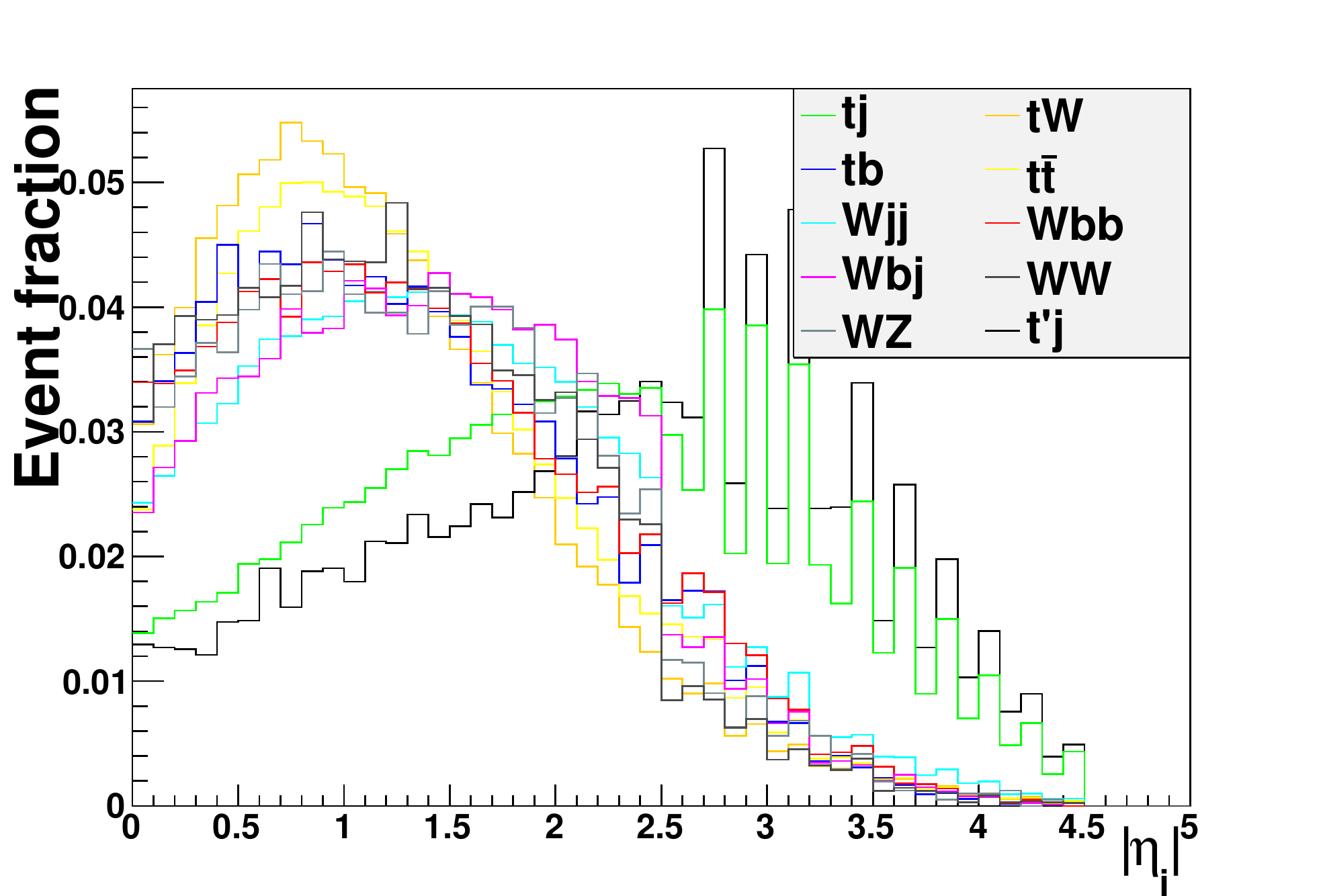}}

 \subfigure[]{
    \label{bW:a}
    \includegraphics[width=0.45\textwidth]{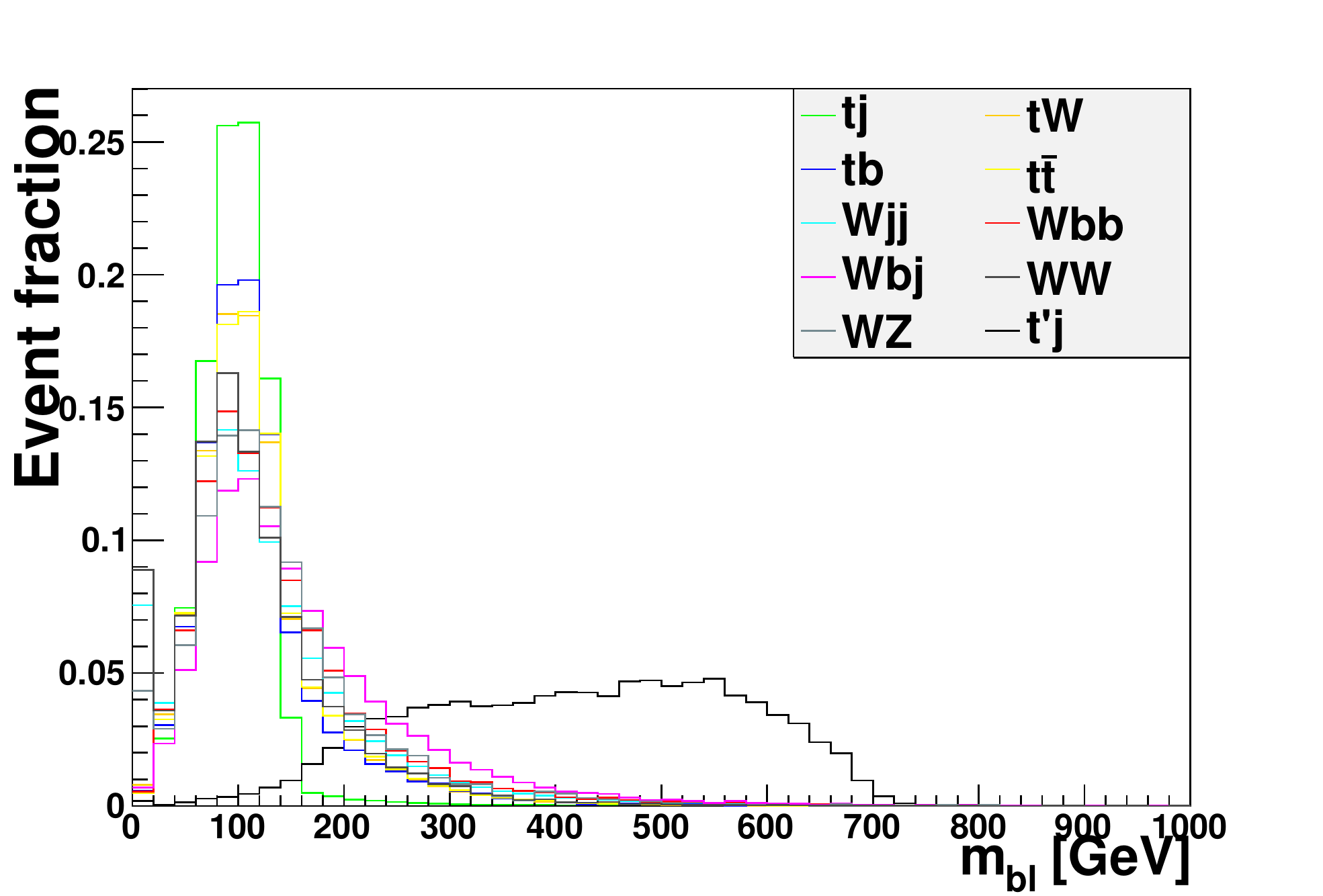}}
\hspace{0.05in}
 \subfigure[]{
    \label{bW:b}
 \includegraphics[width=0.45\textwidth]{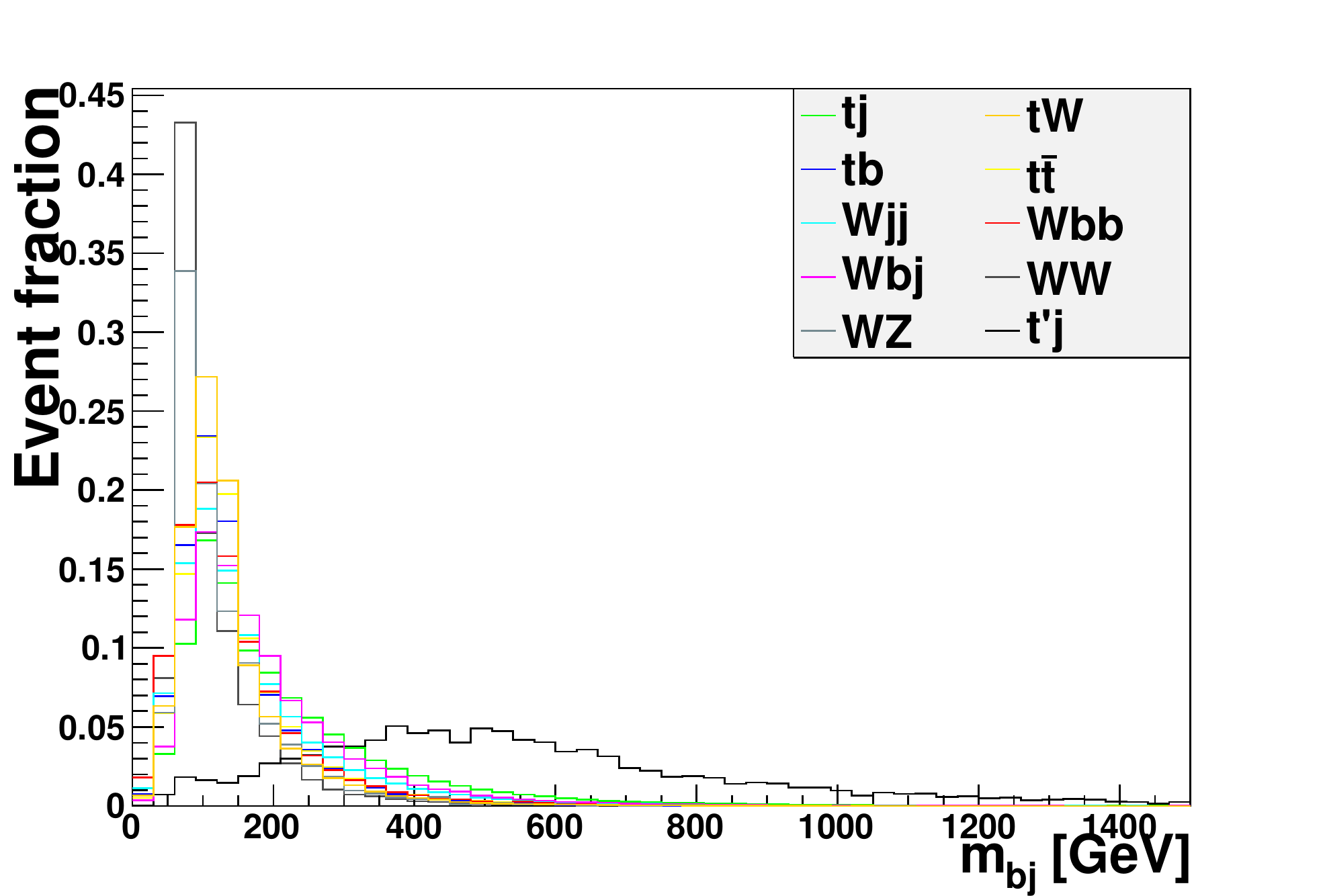}}

  \caption{Distributions for the signal and backgrounds of $(a)H_{T}, (b) p_{T,b}, (c) p_{T,\ell}, (d) |\eta_{j}|,$ $(e) m_{b\ell}, 
  (f) m_{bj}$ after the basic cuts in the $t^\prime \rightarrow bW$ decay channel. The shapes are normalized to unit area.}
  \label{bWdistribution}
\end{figure}

We first impose the following basic cuts to reduce the backgrounds:  
\begin{itemize}
\item{1. There is exactly one isolated lepton.}
\item{2. The missing transverse energy $\slashed E_T$  is required to be larger than 10 GeV.}
\item{3. There are exactly one $b$-tagged jet and no more than three jets in total. If there are two untagged jets, we adopt the one with higher absolute value of $\eta$ as the forward jet candidate.}
\end{itemize}
We do not reconstruct the leptonic $W$, because the reconstruction of the longitudinal momentum of the neutrino from $W$ decay is not good enough due to the large uncertainty in our energy scale. We only take advantage of the large $p_T$ of the b-tagged jet and the lepton decaying from the relatively high mass of $t^\prime$ and the forward nature of the untagged jet from the light quark produced together with the top partner. Based on the kinematical distributions of the signal and backgrounds in Figure~\ref{bWdistribution}, we impose the following cuts furthermore to get a high significance:
\begin{itemize}
\item{4. We require the scalar sum of the transverse momenta of the $b$-tagged jet, the untagged jet, and the lepton  to have $H_{T} >$ 500 GeV. }
\item{5. We require the b-tagged jet to have $p_T >$ 200 GeV, the lepton to have $p_T > $ 150 GeV and the light untagged jet to have $|\eta_j| >$ 2.5.}
\item{6. The invariant mass of the $b$-tagged jet and the lepton is required to have $m_{b\ell} >$ 400 GeV, while the invariant mass of the $b$-tagged jet and the untagged jet $m_{bj}$ is larger than 500 GeV.}
\end{itemize}

We present the cut flow of the signal and background events in Table~\ref{cutflow_bW}, where the second column corresponds to the number of events we generated by Madgraph5 and the third column denotes  the events normalized with the luminosity of 25 $\text{fb}^{-1}$. We finally obtain 11.3 signal and 63 background events in this decay channel, thus getting a local significance of  1.4 $\sigma$ assuming sin$\alpha$ = 0.2. 
We further explore the discovery potential by varying the mass of the $t'$ and the result is shown in Table~\ref{bW0.2} for sin$\alpha$ = 0.2 and Table 
\ref{bW0.3} for sin$\alpha$ = 0.3. As we can observe, the local significance decreased slowly from 3.0 $\sigma$ to 1.8 $\sigma$ by varying the $t'$ mass. This is because of the slowly decreasing cross section and the larger transverse momentum of the objects as $t'$ becomes heavier so that the cut efficient is higher. Using these results, we depict the expected 95\% C. L. exclusion region of this channel  in Figure~\ref{comb95}, from which we can see that this channel will set a strong constraint on  the size of mixing sin$\alpha$  for $m_{t^\prime}$ up to 1 TeV. We also show the constraints for purely chiral fourth generation which has an unit branching fraction of $t^\prime \rightarrow bW$. 

\begin{table}[ht]
\begin{center}

\begin{tabular}{c|cccccc}
     Process          &Generated  &  Normalized&  Cut 1-3 & Cut 4&  Cut 5 & Cut 6\\
  \hline
    $Wjj$                  &39999995 &  99000000 &  473641   &4893 & 87     &47\\
    $t\bar{{t}}$       &10000000  & 5950000   &  311437    &1994 &15     &9.5\\
    $Wb\bar{{b}}$               &    400000  & 288000     & 12884      &234   &2.9      &2.2 \\
    $Wbj$                &  1000000  & 960000    &  37251      &338   &3.8      &1.9   \\
    $WW$                &   2000000 &  1335000  &  8834      &53      &1.3       &1.3\\
    $tW$                  &   999998 &  560000    & 42563      &225    &3.4      & 1.1  \\
  \hline
    $tb$                   &300000   & 140000      &6027      &65      & 0        &0\\
    $tj$                    &2999999  &2200000   &119071   &614   &0.7       &0\\
    $WZ$                 & 500000   &550000     &4505       &19     &0           &0\\
  \hline
   Total background&       &  110983000&1016210&8435  &114 & 63\\
\hline
   $t' j \rightarrow b W j$ &100000&1463&122&52     &16   &11.3\\
  \hline
\end{tabular}
\end{center}
\caption{Cut flows for the signal and backgrounds in the analysis of the single $t^\prime$ production channel with $t^\prime$ decaying 
to $b$ and $W$ boson assuming $m_{t^\prime}$ = 700 GeV and sin$\alpha$ = 0.2. Charge conjugate production modes are implied in the process. Results are shown for the signal assuming the 50\% branching ratio of $t^\prime \rightarrow bW$ decay. }
\label{cutflow_bW}
\end{table}

\begin{table}[ht]
\begin{center}

\begin{tabular}{c|c|c|c|c|c}
$m_{t^\prime}$(GeV) & $S$(no cut)&$S$  & $B$ &  $\frac{S}{\sqrt{S+B}}$& Significance($\sigma$)\\
  \hline
  700      &  1463 &11.3&   63& 1.3 & 1.3 \\
  750      &  1153 & 11.1&   63& 1.3&  1.3\\
  800      &  911 &10.2 &   63& 1.2  &  1.2 \\
  850      &  726  &9.8 &   63 & 1.1  &  1.2 \\
  900      &   581 &8.6&   63& 1.0    &  1.1\\
  950      &   468 &7.0 &   63& 0.8  &   0.8 \\
 1000      &   378 &6.3 &   63&0.8  &   0.7 \\
  \hline
\end{tabular}
\end{center}
\caption{Results for different masses of the top partner assuming sin$\alpha$ = 0.2 in  the $t^\prime \rightarrow bW$ decay channel. }
\label{bW0.2}
\end{table}

\begin{table}[ht]
\begin{center}

\begin{tabular}{c|c|c|c|c|c}
$m_{t^\prime}$(GeV) & $S$(no cut)&$S$  & $B$ &  $\frac{S}{\sqrt{S+B}}$&Significance($\sigma$)\\
  \hline
  700      &  3291 &25.4&   63& 2.7 &3.0\\
  750      &  2593 & 25.0&   63& 2.7&2.9\\
  800      &  2050 &23.0 &   63& 2.5  &2.7\\
  850      &  1634  &22.1 &   63 & 2.4  & 2.6\\
  900      &   1308 &19.4&   63& 2.1 & 2.4\\
  950      &   1052 &15.8 &   63& 1.8  &1.9\\
 1000      &   849 &14.2 &   63&1.6 &1.7\\
  \hline
\end{tabular}
\end{center}
\caption{Results for different masses of the top partner assuming sin$\alpha$ = 0.3 in the $t^\prime \rightarrow bW$ decay channel.}
\label{bW0.3}
\end{table}

\subsection{$th$ decay channel}
\label{th}

Let us now turn to describe the search strategy for the $3bjW$ channel. The Main background comes from the  $t\bar{t}$+jets production with semi-leptonic decay of the top quark pair. This is the most challenging one because of its largest cross section ($\sim$ 238 pb) and large miss-b-tagging rate ($\sim$ 20\%) of the c-quark jets from the hadronically decaying $W$ boson \cite{ATLAS:Tth}.
Other backgrounds like $t\bar{t}b\bar{b}$, $W b \bar{b}$+jets  can be neglected under our consideration.

Parton-level events for the signal and background are generated by MadGraph5~\cite{Alwall:2011uj} and interfaced to PYTHIA6.420~\cite{Sjostrand:2006za} for 
parton-showering and hadronization. For $t\bar{t}$ + jets, MLM matching scheme implemented in pythia is adopted to avoid double-counting in certain regions of phase space. To simulate a realistic experiment environment, we include the  smearing effect by using the tracks and the calorimeter tower information, output by the energy flow algorithm from the Delphes 3.0.5 \cite{Ovyn:2009tx}, as the (fat)-jet constituents. In what follows, we describe the physical object reconstruction techniques and the selection criteria for the candidate events.

Jets are reconstructed using Fastjet 3.0.3~\cite{Cacciari:2011ma}. Firstly, we search for  the Higgs fat jet using Cambridge-Aachen algorithm with radius $R = 1.4$ and require its two leading subjets b-tagged. Once the Higgs fat jet is reconstructed, we erase its constituents from the input particles, while the  remaining ones are clustered into narrow jets using anti-$k_t$ algorithm with a width parameter $R=0.4$. Furthermore, the narrow jets are required to satisfy $|\eta|< 4.5 $ and $p_T > 30$ GeV and only  b-jets satisfying $|\eta|< 2.5$ are considered. In addition, we take the b-tagging efficiency of 70\% for $b$ quark jets, miss-b-tagging probability of 1\% for light jets and 20\% for charm jets \footnote{We notice that Ref. \cite{Vignaroli:2012nf} assume the same light and charm quark mis-b-tagging rate, therefore underestimates the backgrounds.}.  Secondly, leptons are required to have $|\eta|< 2.5 $, $p_T > 25$ GeV and to be isolated. The isolation requirement is the same with that described previously in the $bW$ decay channel.

To reconstruct the boosted Higgs fat jet, we first follow the BDRS~\cite{Butterworth:2008iy} procedure to decompose the fat jets 
which satisfy $p_T >$ 40 GeV to two subjets $j_1$ and $j_2$ with $m_{j_1} > m_{j_2}$. Next, we require a significant mass drop 
condition $m_{j_1} < \mu m_{j}$ with $\mu = 0.667$ and that the splitting is not too asymmetric, i.e., y = min($p_{T,j_1}^2,p_{T,j_2}^2)\Delta R^2_{j_1,j_2}/m_j^2 > y_{\text{cut}} $ with $y_{\text{cut}} = 0.09$. Finally, we filter the Higgs neighbourhood similar to the BDRS, 
resolving the fat jets on a finer angular scale, $R_{\text{filt}} = \text{min}(0.35, R_{j_1j_2}/2)$ and taking the three hardest 
objects (subjets) that appear, which will eliminate much of the underlying event contamination. 

\begin{figure}[ht]
 \centering
 \subfigure[]{
    \label{th:a}
    \includegraphics[width=0.48\textwidth]{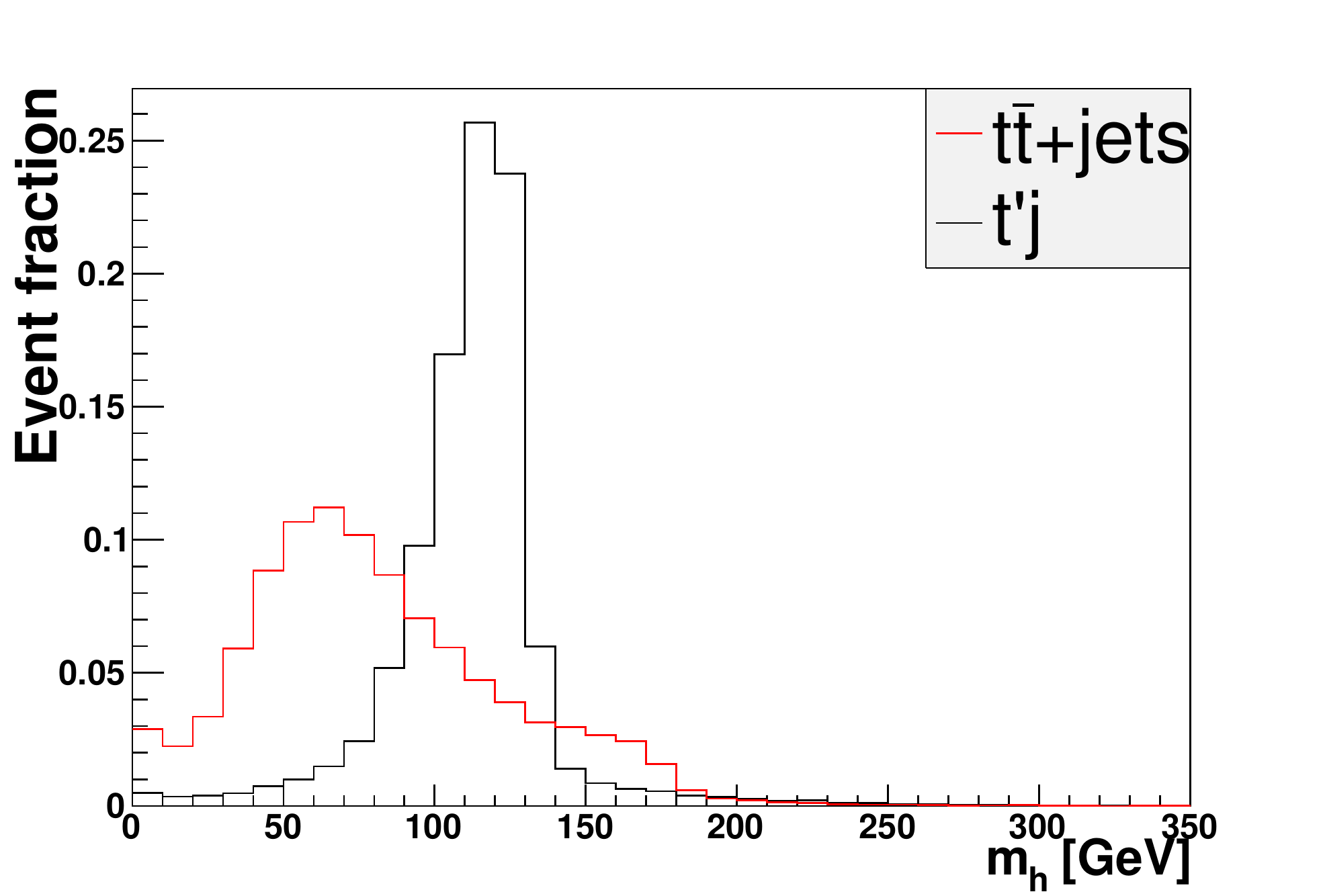}}
\hspace{0.01in}
 \subfigure[]{
    \label{th:b}
 \includegraphics[width=0.48\textwidth]{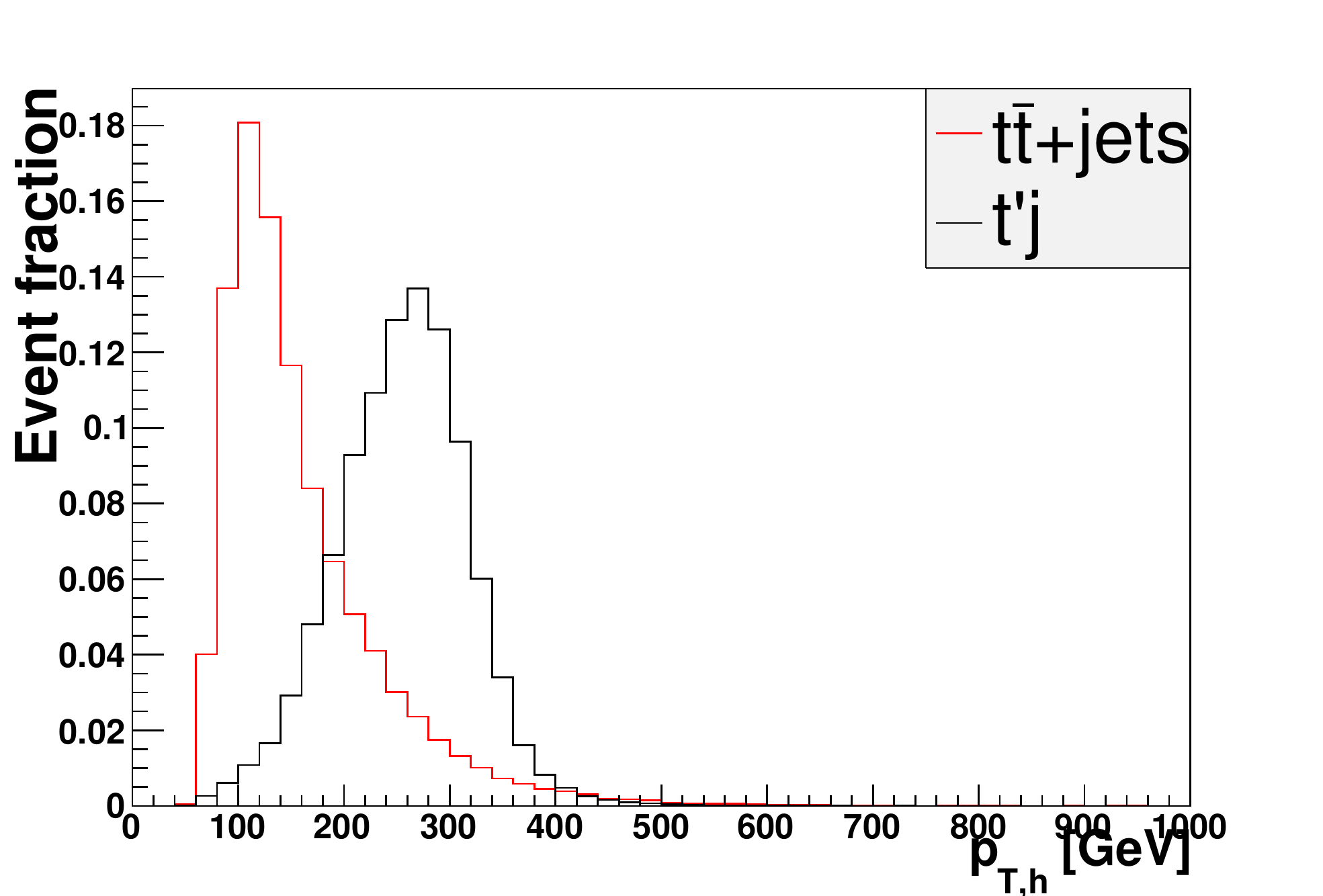}}
 \subfigure[]{
    \label{th:a}
    \includegraphics[width=0.48\textwidth]{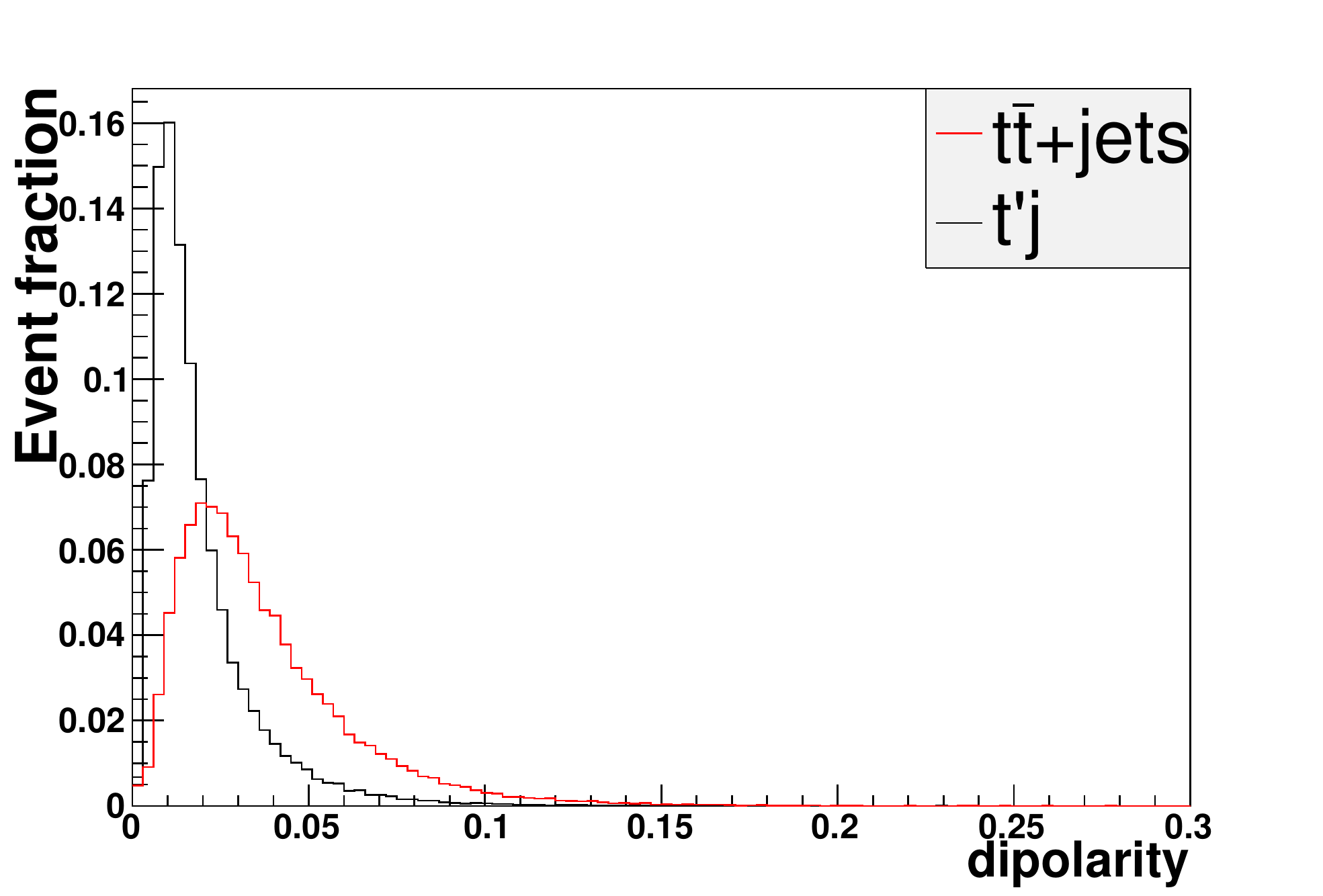}}
\hspace{0.01in}
 \subfigure[]{
    \label{th:b}
 \includegraphics[width=0.48\textwidth]{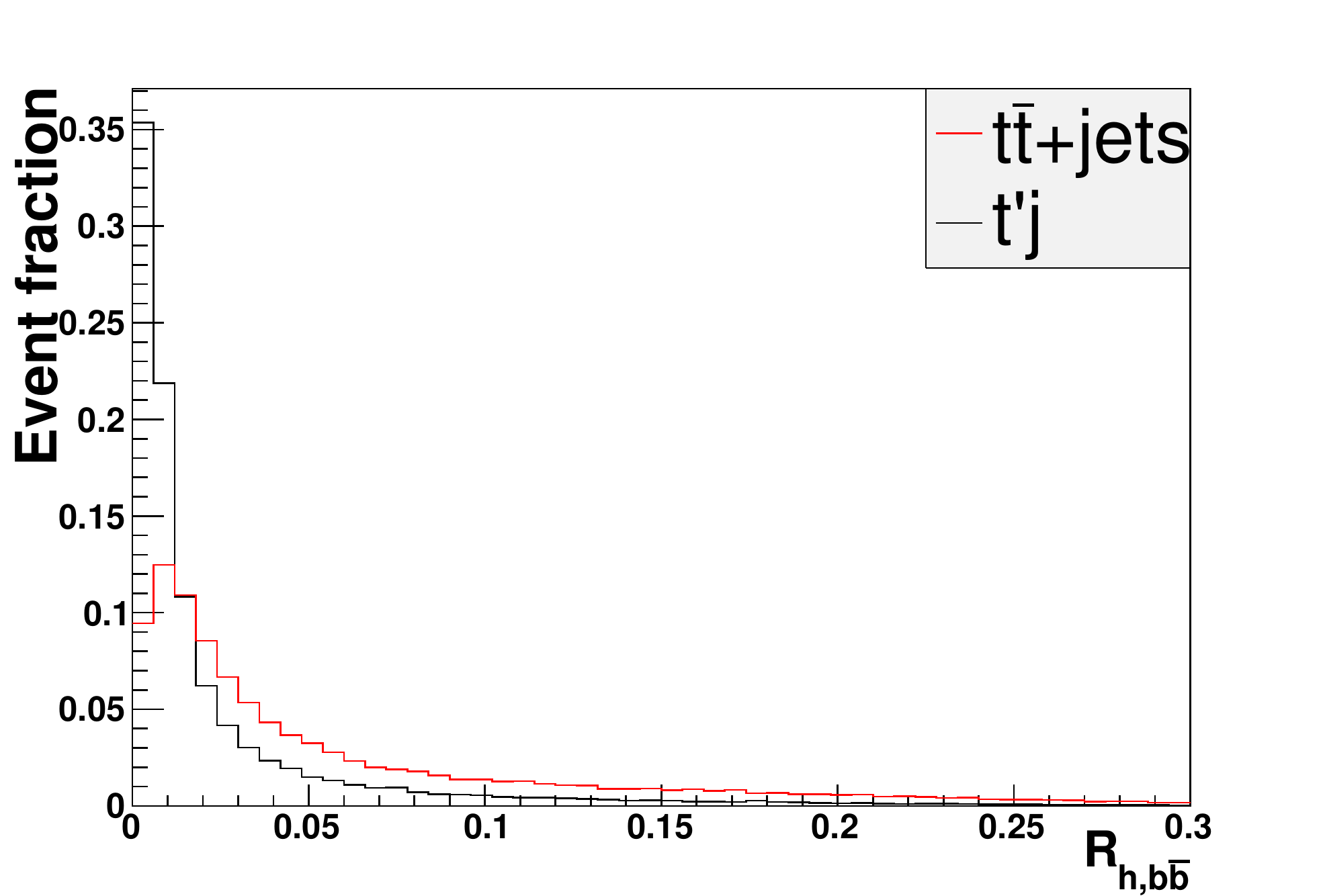}}
 \subfigure[]{
    \label{th:a}
    \includegraphics[width=0.5\textwidth]{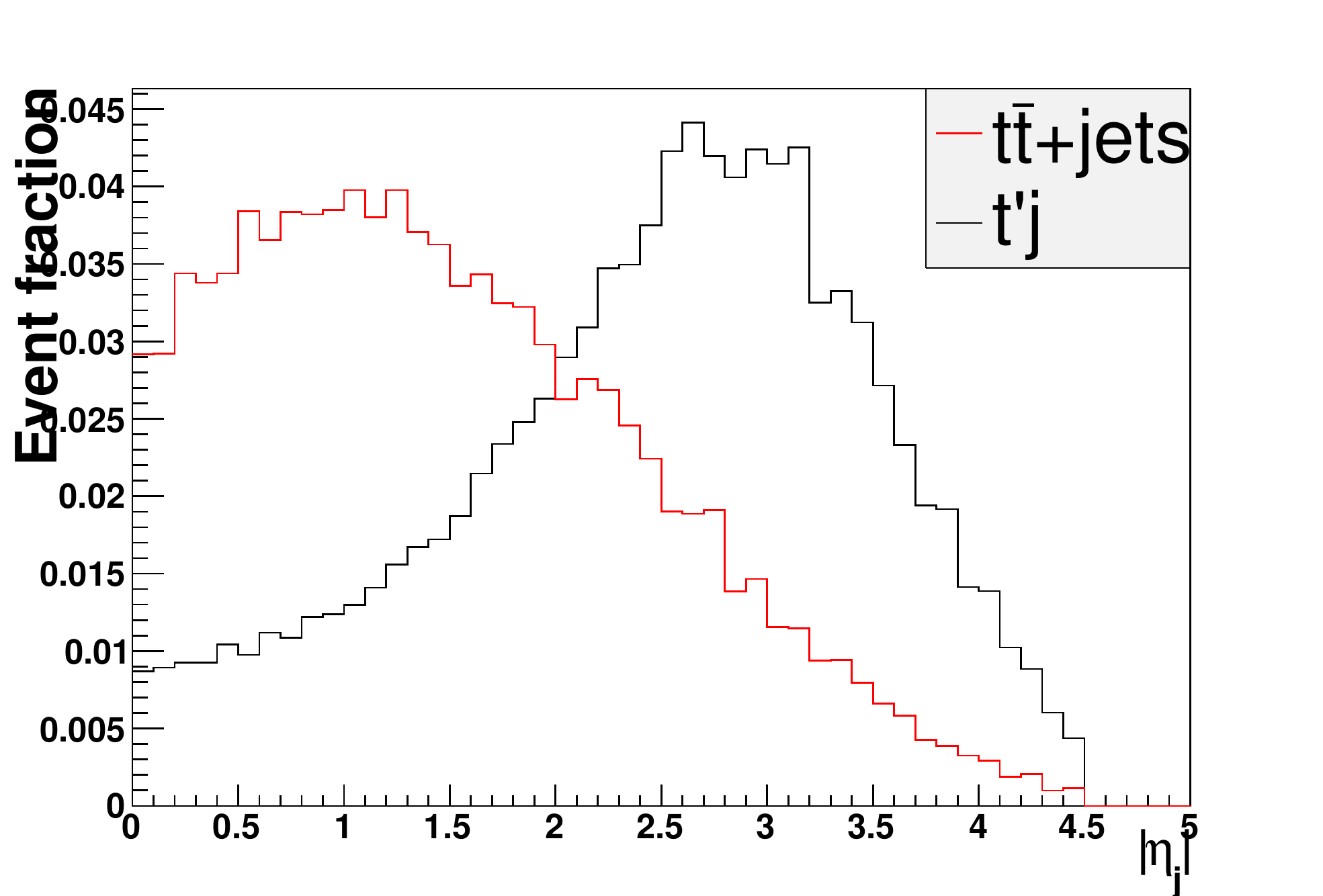}}
 
  \caption{Distributions for the signal and background of $(a) m_h$, $(b) p_{T,h}$, $(c) \text{dipolarity}$, $(d) R_{h,b\bar{b}}$, $(e) |\eta_{j}|$ after the basic cuts in the $t^\prime \rightarrow th$ decay  channel. The shapes are normalized to unit area.}
  \label{thdistribution1}
\end{figure}

\begin{figure}[ht]
 \centering
 \subfigure[]{
    \label{th:f}
    \includegraphics[width=0.48\textwidth]{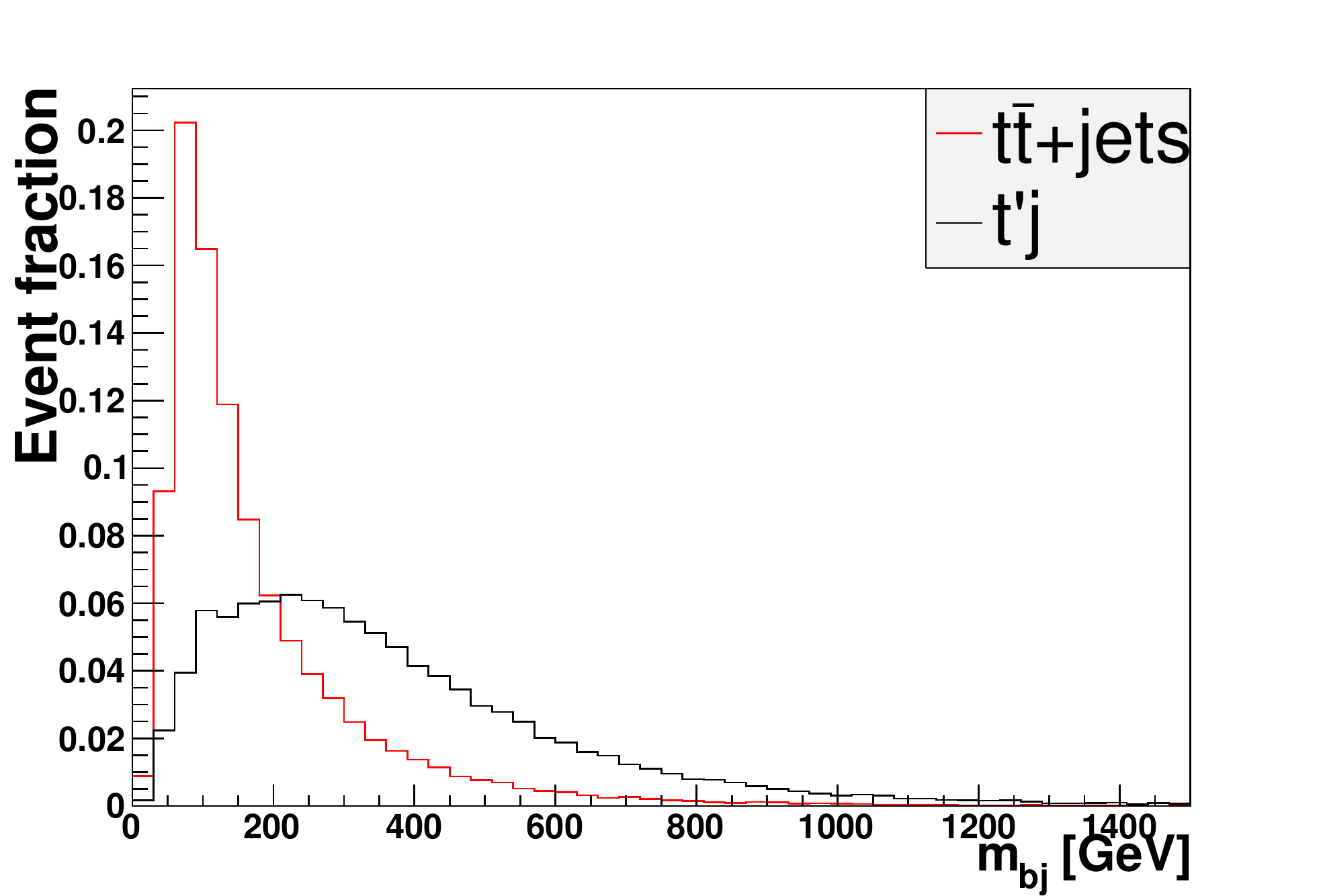}}
 \hspace{0.05in}
 \subfigure[]{
    \label{th:g}
 \includegraphics[width=0.48\textwidth]{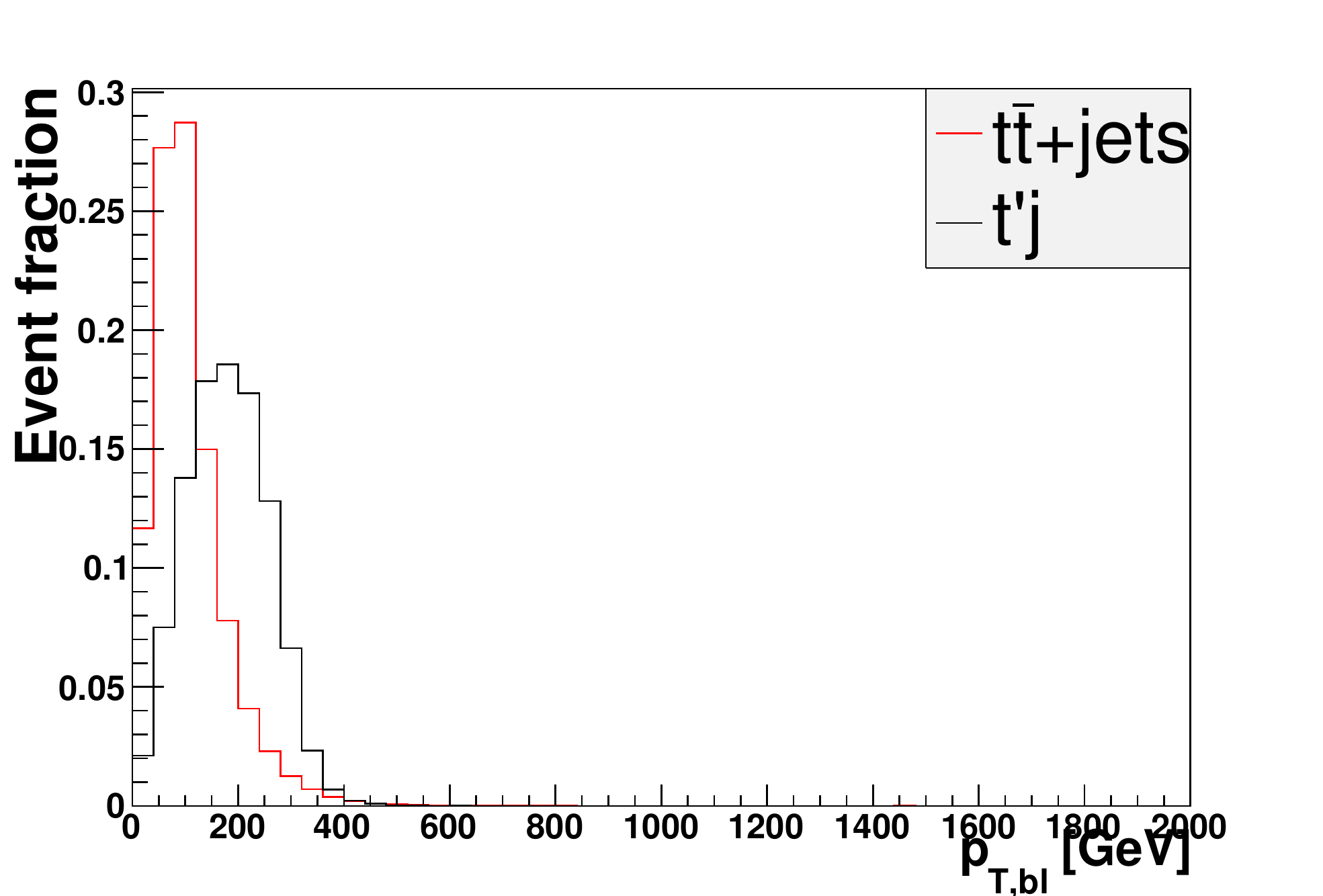}}
 \subfigure[]{
    \label{th:h}
    \includegraphics[width=0.48\textwidth]{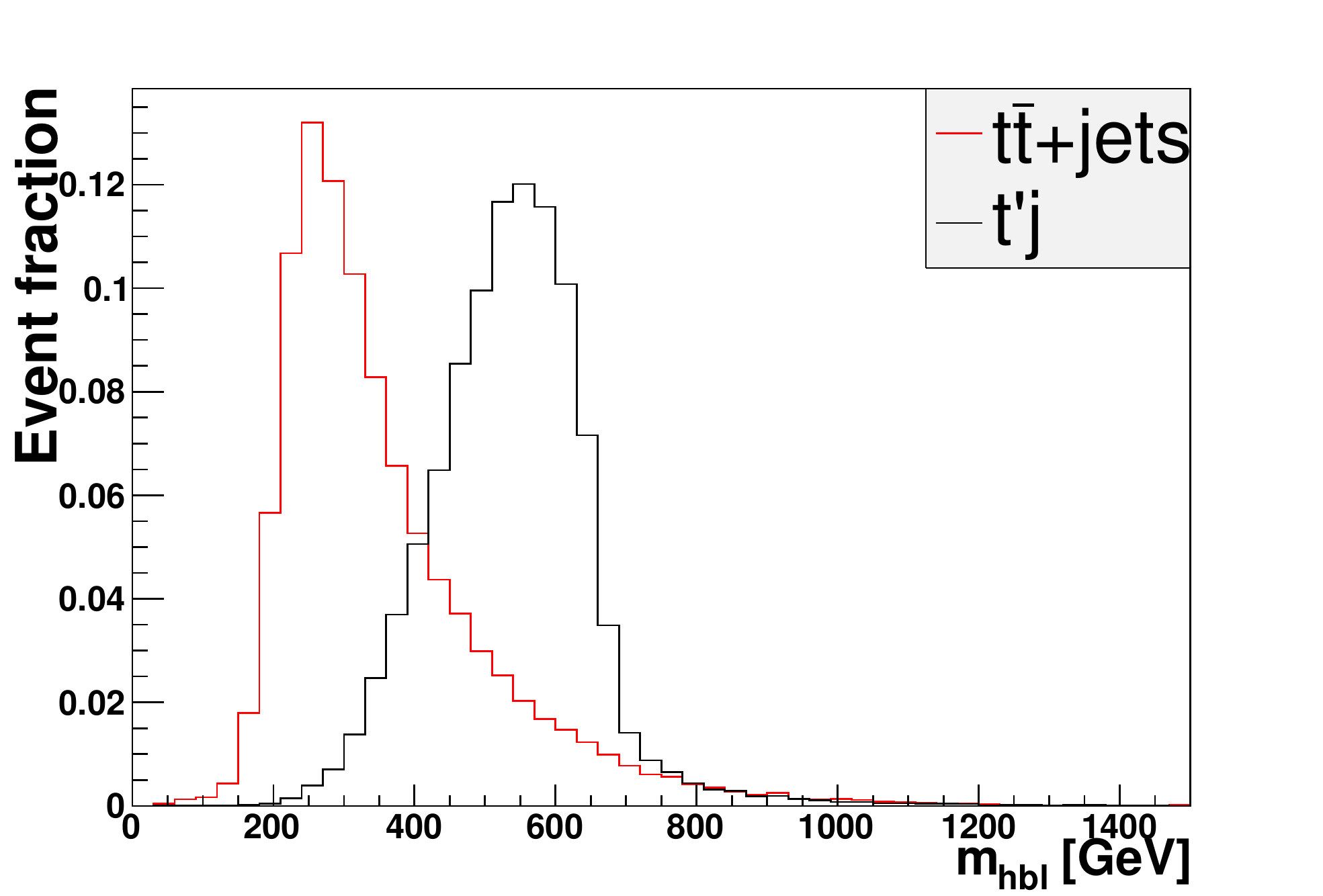}}
\hspace{0.05in}
 \subfigure[]{
    \label{th:i}
 \includegraphics[width=0.48\textwidth]{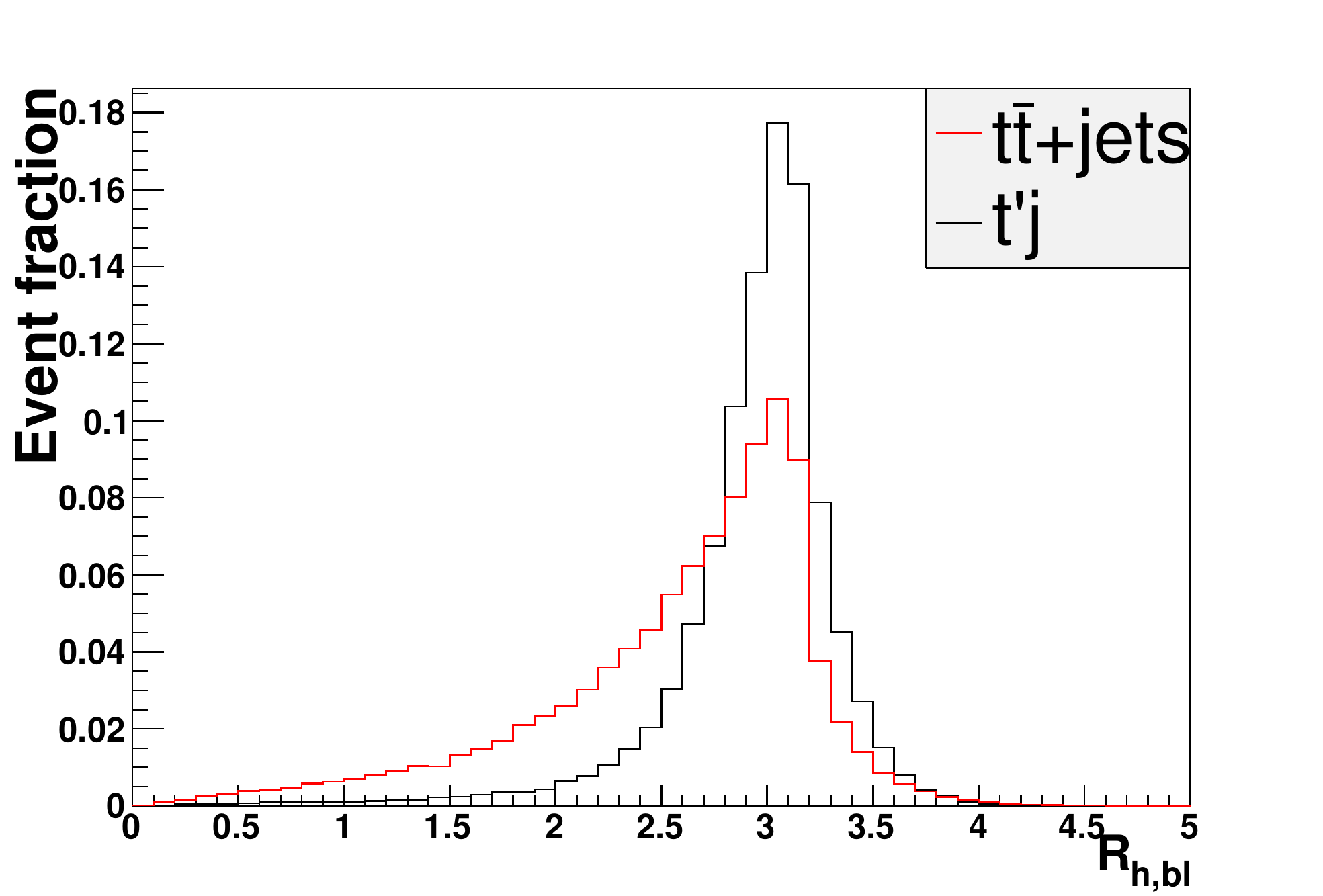}}
 
  \caption{Distributions for the signal and background of $(a) m_{bj}$, $(b) p_{T,b\ell}$, $(c) m_{h,b\ell}$, $(d) R_{h,b\ell}$ after the basic cuts in the $t^\prime \rightarrow th$ decay  channel. The shapes are normalized to unit area.}
  \label{thdistribution2}
\end{figure}

Since Higgs is a color singlet, the $b$-partons from its decay are color-connected and most radiations will be contained within two angular cones around the $b$-partons as a result of angular ordering. This radiation pattern controlled by the color flow will provide complimentary information, which is used to define the $pull$ for discrimination~\cite{Gallicchio:2010sw}. In our case, the Higgs is moderately boosted and it may be more useful to use the jet dipolarity instead~\cite{Hook:2011cq} \footnote{We thank Tao Liu for the useful discussions and the feedback of their cutting efficiency from the pull angle in Ref. \cite{Berenstein:2012fc}}. The dipolarity $\mathcal{D}$ is defined as: 
\beq
\mathcal{D} \equiv \frac{1}{R_{12}^2}\sum_{i\in J} \frac{p_{Ti}}{p_{TJ}}R_i^2, 
\eeq
where $R_{12}$ is the angular distance between the two subjets, $R_{12}^2 = (\eta_1 - \eta_2)^2 + (\phi_1 - \phi_2)^2$, and $R_i$ is the minimum euclidean distance between each particle  $(\eta_i,\phi_i)$ in the fat jet and the line segment that joins $(\eta_1,\phi_1)$ and $(\eta_2,\phi_2)$ in the $\eta-\phi$ plane. In our calculation, $(\eta_1,\phi_1)$ and $(\eta_2,\phi_2)$ are identified with those of  the two $b$-tagged subjects in the filtered Higgs fat jet and we include all the radiations contained within the two cones of size $R_{12}/\sqrt{2}$ centered around the $b$-tagged subjects, i.e. we also include the soft radiations which are discarded by the mass drop criterion and filtering. We do not simulate underlying events, but we expect this will not change our results significantly because we choose the smaller cone when including the soft radiations.

The event selection requires at least one filtered fat jet with two leading subjets b-tagged and the highest $p_T$ one will be referred as the $h \rightarrow b\bar{b}$ candidate. Other fat jets will be dropped and the particles will be used to reconstruct the narrow jets with $R = 0.4$ as illustrated above. We then impose the following basic cuts to reduce the backgrounds:  
\begin{itemize}
\item{1. There is exactly one isolated lepton. }
\item{2. The missing transverse energy $\slashed E_T$  is required to be larger than 10 GeV.}
\item{3. There is at least one $b$-tagged jet and one untagged jet in addition to the Higgs fat jet.
The leading $b$-jet will be referred to the top decaying $b$-jet candidate and the untagged one 
with the largest absolute value of $\eta$ will be treated as the light jet produced in association with the top partner.}
\end{itemize}

From Table~\ref{cutflow_th}, we can see that even after the basic cuts, the backgrounds are significantly rejected. For further analysis, we show the distributions of some important kinematical variables  after the basic cuts in Figure~\ref{thdistribution1} and Figure~\ref{thdistribution2}, from which we can infer that one can obtain good discrimination between signal and backgrounds. Here, several remarks are in order. Firstly,  the distribution of the invariant mass of the reconstructed boosted Higgs fat jet has been broadened as a result  of  the limited resolution of the detector. Because of the small number ($\sim$ 34) of signal events in hand after the basic cuts, we impose a relatively large mass window around the true Higgs mass.
 Secondly, since the reconstruction of the leptonically decaying $W$ boson is not as good as expected ($\sim$ 30\%), we just abandon the information of the reconstructed top and turn to the system of the leading $b$-tagged jet and the unique isolated lepton. Some kinematical variables are found to be very useful such as the transverse momentum of the ($b,\ell$) system $p_{T,b\ell}$, the invariant mass of the Higgs fat jet, the $b$-tagged jet and the lepton $m_{h,b\ell}$ and the distance of the Higgs jet with the ($b,\ell$) system $R_{h,b\ell}$. They tend to have larger value for the signal as a result of the heavy $t'$ that we search for. In order to enhance the significance, we in further impose the following conditions:

\begin{itemize}
\item{4. We require the Higgs fat jet  to have $p_{T,h} >$ 200 GeV and  the distance between the Higgs and its $b\bar{b}$ subsystems $R_{h,b\bar{b}}$ 
is smaller than 0.05. We also impose a relatively large mass window for it around the true Higgs mass, $m_h \in [100,130]$ GeV.  In addition, we require the dipolarity of the Higgs fat jet is smaller than 0.02.}
\item{5. The light untagged jet is required to have $|\eta_j| >$ 2.5 and the invariant mass of the b-tagged jet and the untagged jet $m_{bj}$ is larger than 200 GeV.  }
\item{6.  We choose $p_{T,b\ell} >$ 160 GeV, 450 GeV $< m_{h,b\ell} <$ 650 GeV and the distance of  Higgs candidate with  the system of the leading b-tagged jet and the lepton $R_{h,b\ell}$ is larger than 2.0.}
\end{itemize}

We present the cut flow of the signal and background events in Table~\ref{cutflow_th} assuming
sin$\alpha$  = 0.4, where
the second row corresponds to the number of events we generated by Madgraph5 and the third row denotes the events normalized with the luminosity of 25 $\text{fb}^{-1}$.
We obtained 4.1 signal and 6.6 background events with $S/\sqrt{S+B} \sim $ 1.3, and a local significance of 1.5 $\sigma$. In order to gain more discrimination power, we apply the boosted decision tree (BDT) method~\cite{BDT} which are implemented in 
the ROOT TMVA package~\cite{Hocker}.  In addition to the kinematical variables described above, we add the following: $p_{T,b}, p_{T,j}, p_{T,\ell}, \eta_{b}, m_{b\ell}, m_{T,W}$, where $m_{T,W}$ is the transverse mass of the leptonically decaying W boson, which is defined as $m_{T,W} = \sqrt{2p_{T,\ell}\slashed E_T (1-\text{cos}\Delta\phi_{\ell,\slashed E_T})}$. \\

\begin{figure}[ht]
\begin{center}
  \includegraphics[width=0.47 \textwidth]{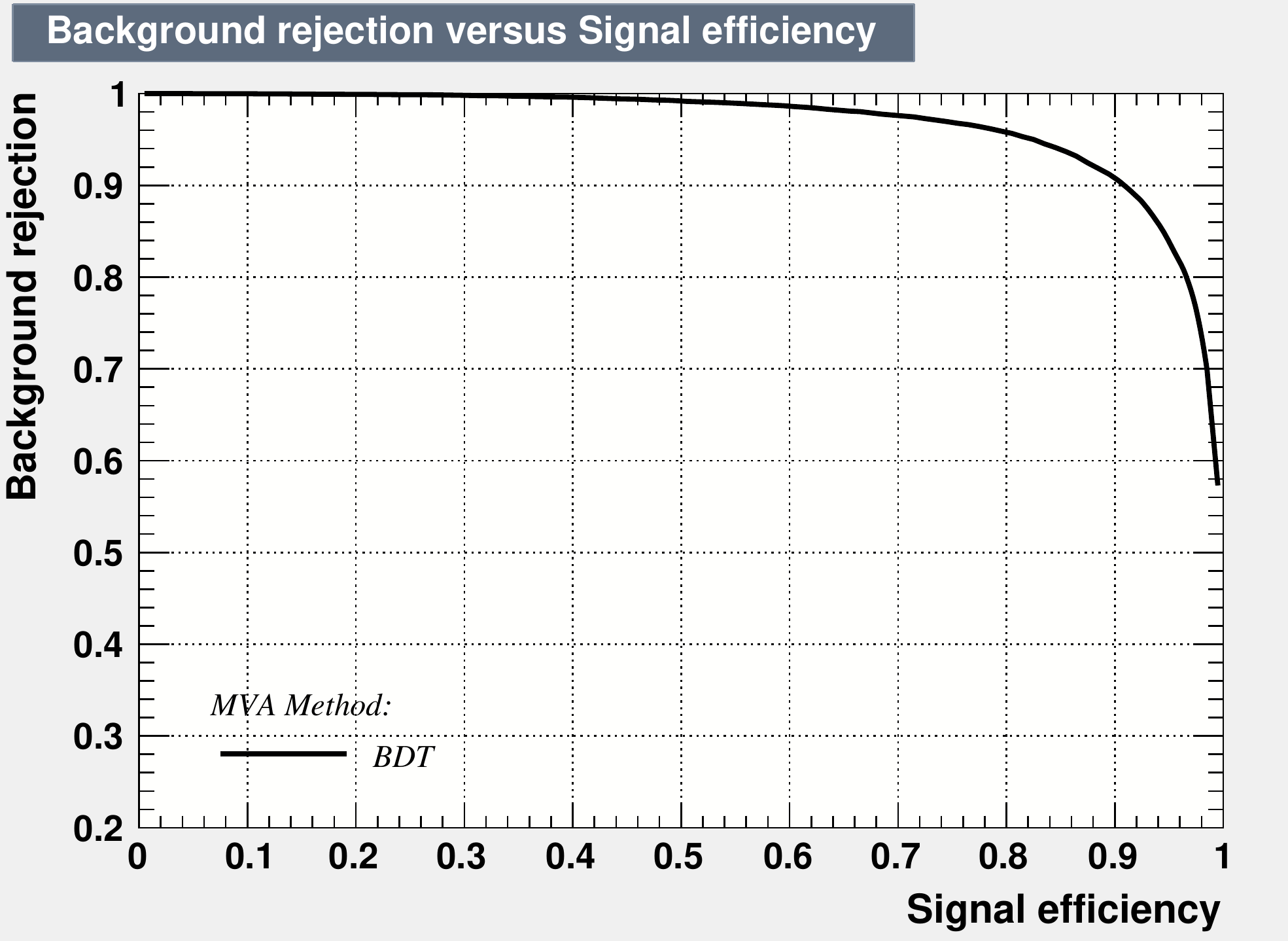} \qquad
  \includegraphics[width=0.47 \textwidth]{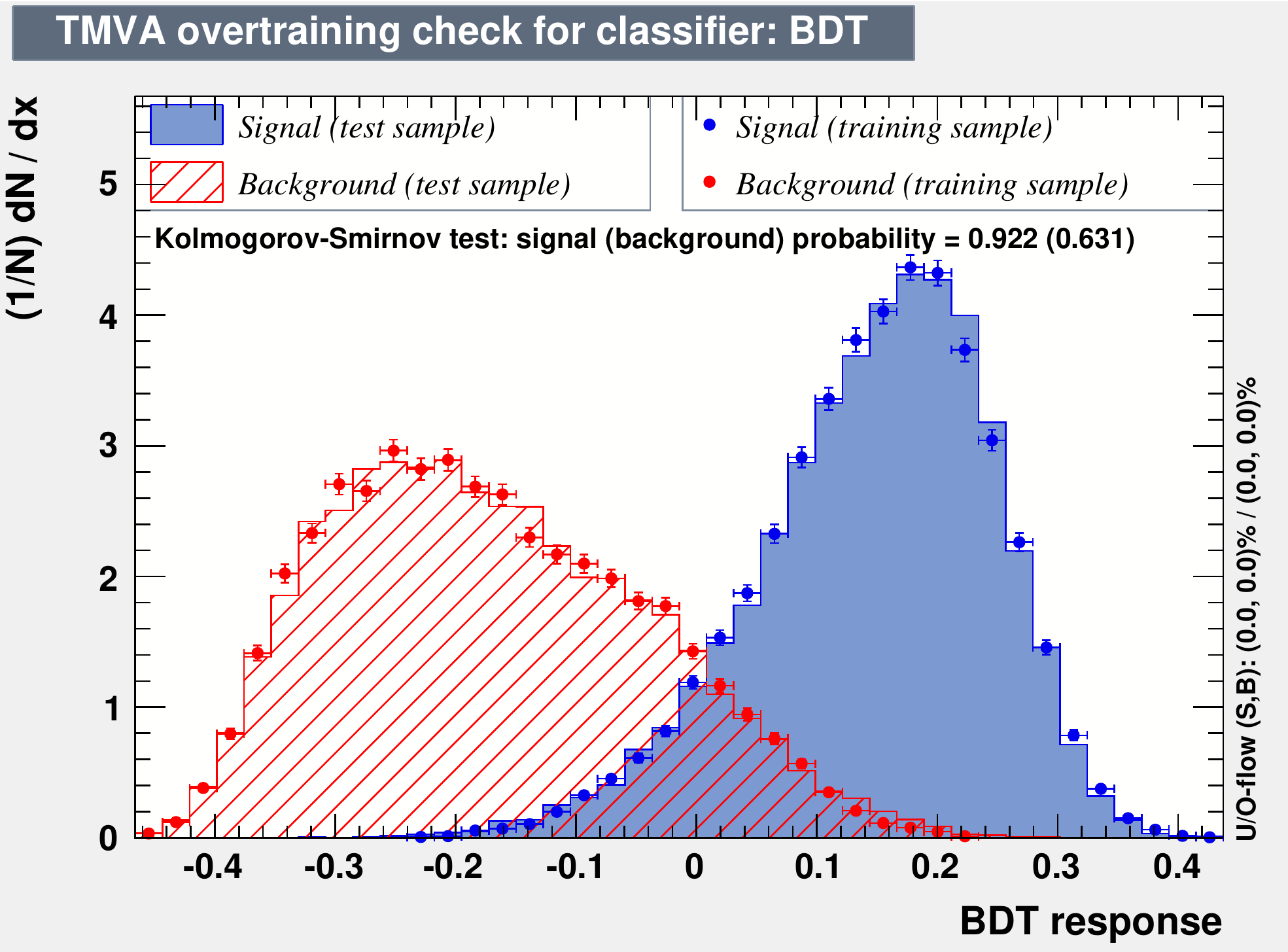}\\
\end{center}
  \caption{The outputs of BDT analysis. Left panel: background rejection vs. signal efficiency. Right panel: Normalized distributions of BDT response for signal and background. }
  \label{BDT}
\end{figure}

\begin{figure}[ht]
\begin{center}
  \includegraphics[width=0.7\textwidth]{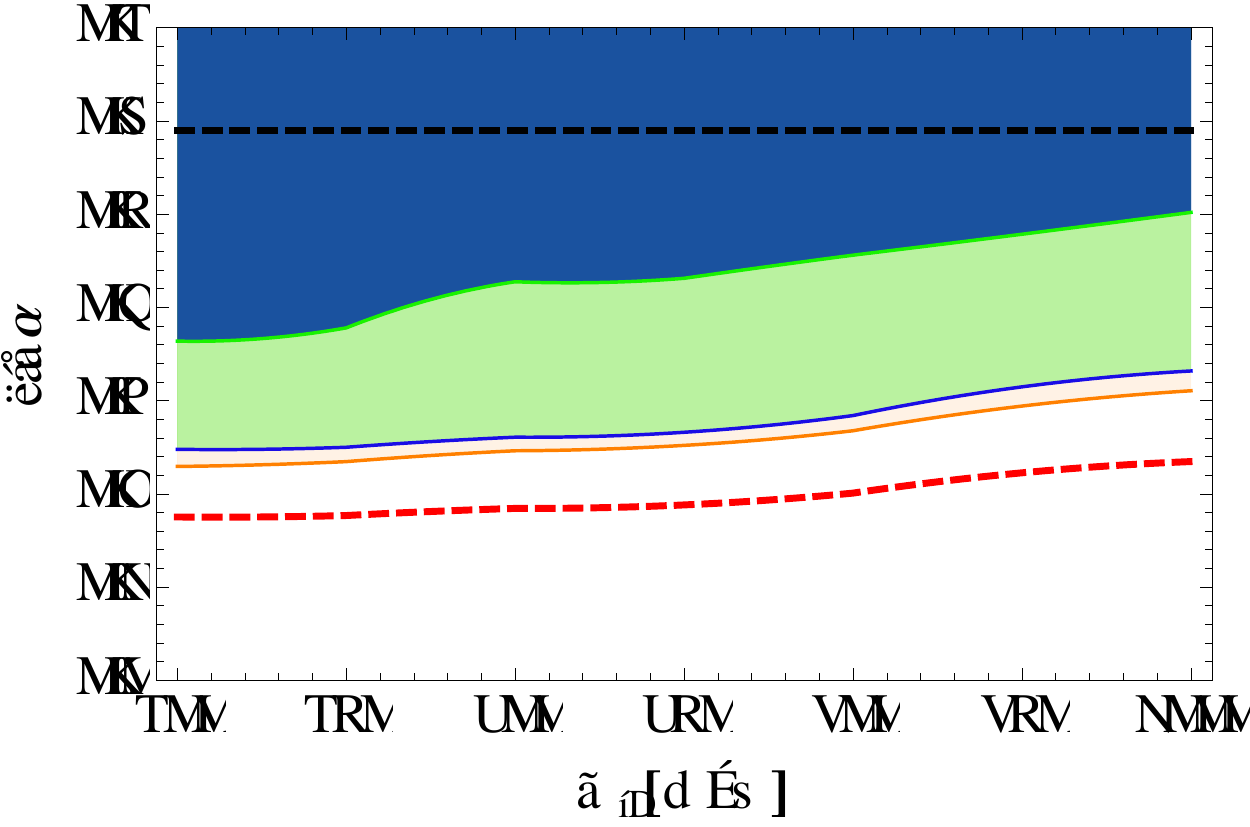}
\end{center}
  \caption{Expected 95\% C. L. Exclusion region in the $m_{t^\prime}$-sin$\alpha$ plane. Green region: $t^\prime \rightarrow$ bW decay channel; Blue region: $t^\prime \rightarrow$ th decay channel; orange region: combination of the two channels. The black dashed line is the bound from the current $V_{tb}$ constrain and red dashed line is the constrain for the purely chiral fourth generation. }
  \label{comb95}
\end{figure}

We have trained 1000 decison trees and the outputs are presented in Figure~\ref{BDT}, from which we observe that good discrimination can be obtained between signal and background. We finally obtain 9.7 signal and 10.1 background events, thus getting S/$\sqrt{S+B} \sim $ 2.2, and a local significance of 2.7 $\sigma$. In Table~\ref{th0.4}, we show the BDT results for different mass of $t^\prime$ with sin$\alpha$ = 0.4, from which we can see that this channel can have evidence up to 1 TeV. Using these results, we plot the expected 95$\%$ C. L. exclusion region in the $m_{t^\prime}$-sin$\alpha$ plane in Figure~\ref{comb95}.  In this figure, we also show  the combined 95\% C. L. exclusion region of $t^\prime \rightarrow th$ and $t^\prime \rightarrow bW$ decay channels, from which we can observe that the  single $t^\prime$ production channel at the 8 TeV LHC will set a strong constraint on the $t^\prime bW$ coupling with the mass of $t^\prime$ up to 1 TeV. The current $V_{tb}$ constraint from the single top measurements~\cite{CMS:singletop, ATLAS:singletop} only sets a much weaker upper limit on the parameter $\text{sin}\alpha < 0.59$ in our simplified model, which we depict in Figure~\ref{comb95}. 

\begin{table}[ht]
\begin{center}

\begin{tabular}{c|ccc}
                 &$t \bar{{t}}$+jets & $t^\prime j + \bar{t}^\prime j \rightarrow 3b W j  $  \\
  \hline
  Generated      &30732326    & 1999998 \\
  Normalized     & 5950000    & 1697 \\
  Cut 1-3          &    7277       &  35  \\
  Cut 4             &      178         &  15   \\
  Cut 5             &      25.4           &    7.3 \\
  Cut 6             &       6.6           &     4.1  \\
  \hline
   Acceptance    &0.0000011  & 0.0024 \\
  \hline
\end{tabular}
\end{center}
\caption{Cut flows for the signal and background in the analysis of the single $t^\prime$ production channel with $t^\prime$ decaying 
to top and Higgs for sin$\alpha$ = 0.4 and $m_{t^\prime}$ = 700 GeV. Results are shown for  the signal assuming the 25\% branching ratio of $t^\prime \rightarrow t h$ decay.}
\label{cutflow_th}
\end{table}

\begin{table}[ht]
\begin{center}

\begin{tabular}{c|c|c|c|c|c}
$m_{t^\prime}$(GeV) &S(no cut) &S  & B &  $\frac{S}{\sqrt{S+B}}$&Significance ($\sigma$)\\
  \hline
  700      &  1697    & 9.7        &    10.1    & 2.2      & 2.7 \\
  750      &  1337    &10.1         &   14.3     & 2.0    & 2.5  \\
  800      &  1057    &10.5          &   24.4     & 1.8    & 2.0\\
  850      &   842     &9.9           &    25.2    & 1.7    &  1.8 \\
  900      &   674     &7.5           &    15.9    & 1.6    &  1.8\\
  950      &   542     &8.3           &     25.9     & 1.4    & 1.5\\
 1000      &   438    &5.6           &     13.2      & 1.3   & 1.4\\
  \hline
\end{tabular}
\end{center}
\caption{Results for different masses of the top partner assuming sin$\alpha$ = 0.4 obtained by BDT analysis in the $t^\prime \rightarrow th$ decay channel.}
\label{th0.4}
\end{table}

\section{Summary}
\label{sec:conclusion}
In this paper, we have studied the prospects of observing the single $t^\prime$ production at the 8 TeV LHC in the $bW$ and $th$ decay channels with 25 fb$^{-1}$ of integrated luminosity. To illustrate our result, we adopt the simplified model based on the minimal coset $SO(5)/SO(4)$, where the top partner is from the singlet of the unbroken $SO(4)$. We focus on the relatively high mass region of the top partner so that the single production is more efficient than the pair production at 8 TeV LHC. Since the single $t'$ production depends on the $t^\prime$bW coupling and the $t'$ mass (The decay branching ratios are close to the equivalence limit for heavy $t'$), we constrain the parameter space in the $m_{t^\prime}-\text{sin}\alpha$ plane, where $\alpha$ is the left-hand mixing angle between the top quark and the top partner. In the $bW$ decay channel, we rely on the large $p_T$ of  the b-jet, the lepton, and the forward nature of the light jet to suppress the backgrounds. In the $th$ decay channel, we have exploited jet substructure method to tag the boosted Higgs boson with decaying into $b\bar{b}$, where the dominate background is the $t \bar{t}$ + jets with large mis-b-tagging  rate of c-quark from the $W$ decay. We also analyse the jet dipolarity of the Higgs which is a color singlet to improve discrimination ability. Combing the two results from the above two channels, we finally derive the expected 95\% C. L. exclusion region in the $m_{t^\prime}-\text{sin}\alpha$ plane, from which we conclude that the single $t^\prime$ production will set a strong constraint on the $t^\prime$bW coupling ($\sin \alpha \subset [0.2, 0.3]$) at the 8 TeV LHC for  $m_{t'} \subset [700, 1000]$ GeV. Interestingly, our constraint does not sensitively depends on the $t'$ mass because our cut efficiencies are much better for large $t'$ mass. 

The single $t'$ production, which only exists after EWSB through $t -t'$ mixing, is intrinsically connected to the origin of EWSB. The size of such a mixing ($\sin \alpha$), which represents the compositeness of top quark, is a critical parameter to induce the radiative Higgs potential from top quark loops, which give us the most important contribution for EWSB. 
Therefore, we expect our constraints from the single $t'$ production at the early LHC, especially those on the mixing size of a heavy $t'$, 
when elaborated by the experimentalists in the future, would provide much more sophisticated understandings on the true nature of composite EWSB.     

\section*{Acknowledgements}

We would like to thank Kaoru Hagiwara, Qiang Li, Zhao Li, Tao Liu, Martin Jankowiak, Mihoko M. Nojiri, Michihisa Takeuchi, Riccardo Torre, Helge Voss and Wei Zou for helpful discussion. Da Liu is supported by the National Natural Science Foundation of China (Grant Nos. 11275247 and 10821504).



\begin{thebibliography}{99}

\bibitem{Erler:2010sk} 
  J.~Erler and P.~Langacker,
  Phys.\ Rev.\ Lett.\  {\bf 105}, 031801 (2010)
  [arXiv:1003.3211 [hep-ph]].
  
\bibitem{Eberhardt:2010bm} 
  O.~Eberhardt, A.~Lenz and J.~Rohrwild,
  Phys.\ Rev.\ D {\bf 82}, 095006 (2010)
  [arXiv:1005.3505 [hep-ph]].

\bibitem{Murayama:2010xb} 
  H.~Murayama, V.~Rentala, J.~Shu and T.~T.~Yanagida,
  Phys.\ Lett.\ B {\bf 705}, 208 (2011)
  [arXiv:1012.0338 [hep-ph]].
  
\bibitem{Kribs:2007nz} 
  G.~D.~Kribs, T.~Plehn, M.~Spannowsky and T.~M.~P.~Tait,
  Phys.\ Rev.\ D {\bf 76}, 075016 (2007)
  [arXiv:0706.3718 [hep-ph]].
  
\bibitem{Keung:2011zc} 
  W.~-Y.~Keung and P.~Schwaller,
  JHEP {\bf 1106}, 054 (2011)
  [arXiv:1103.3765 [hep-ph]].

\bibitem{Carpenter:2011wb} 
  L.~M.~Carpenter,
  arXiv:1110.4895 [hep-ph].
  
\bibitem{Eberhardt:2012ck} 
  O.~Eberhardt, A.~Lenz, A.~Menzel, U.~Nierste and M.~Wiebusch,
  Phys.\ Rev.\ D {\bf 86}, 074014 (2012)
  [arXiv:1207.0438 [hep-ph]].
  
\bibitem{Eberhardt:2012gv} 
  O.~Eberhardt, G.~Herbert, H.~Lacker, A.~Lenz, A.~Menzel, U.~Nierste and M.~Wiebusch,
  Phys.\ Rev.\ Lett.\  {\bf 109}, 241802 (2012)
  [arXiv:1209.1101 [hep-ph]].
  

\bibitem{Matsedonskyi:2012ym}
  O.~Matsedonskyi, G.~Panico and A.~Wulzer,
  JHEP {\bf 1301} (2013) 164
  [arXiv:1204.6333 [hep-ph]].

\bibitem{Redi:2012ha}
  M.~Redi and A.~Tesi,
  JHEP {\bf 1210} (2012) 166
  [arXiv:1205.0232 [hep-ph]].

\bibitem{Marzocca:2012zn}
  D.~Marzocca, M.~Serone and J.~Shu,
  JHEP {\bf 1208} (2012) 013
  [arXiv:1205.0770 [hep-ph]].
  
\bibitem{Pomarol:2012qf} 
  A.~Pomarol and F.~Riva,
  JHEP {\bf 1208}, 135 (2012)
  [arXiv:1205.6434 [hep-ph]].
 
  
\bibitem{Berger:2012ec} 
  J.~Berger, J.~Hubisz and M.~Perelstein,
  JHEP {\bf 1207}, 016 (2012)
  [arXiv:1205.0013 [hep-ph]].
   
\bibitem{Panico:2012uw} 
  G.~Panico, M.~Redi, A.~Tesi and A.~Wulzer,
  JHEP {\bf 1303}, 051 (2013)
  [arXiv:1210.7114 [hep-ph]].
  
\bibitem{Pappadopulo:2013vca} 
  D.~Pappadopulo, A.~Thamm and R.~Torre,
  arXiv:1303.3062 [hep-ph].

\bibitem{Endo:2011xq} 
  M.~Endo, K.~Hamaguchi, S.~Iwamoto and N.~Yokozaki,
  Phys.\ Rev.\ D {\bf 85}, 095012 (2012)
  [arXiv:1112.5653 [hep-ph]].

\bibitem{Vecchi:2013bja} 
  L.~Vecchi,
  arXiv:1304.4579 [hep-ph].
 

\bibitem{Chatrchyan:2012vu}
  S.~Chatrchyan {\it et al.}  [CMS Collaboration],
  Phys.\ Lett.\ B {\bf 718} (2012) 307
  [arXiv:1209.0471 [hep-ex]].

\bibitem{Chatrchyan:2012af}
  S.~Chatrchyan {\it et al.}  [CMS Collaboration],
  JHEP {\bf 01} (2013) 154
  [arXiv:1210.7471 [hep-ex]].

\bibitem{ATLAS:2012qe}
  G.~Aad {\it et al.}  [ATLAS Collaboration],
  Phys.\ Lett.\ B {\bf 718} (2013) 1284
  [arXiv:1210.5468 [hep-ex]].
 
\bibitem{Harigaya:2012ir}
  K.~Harigaya, S.~Matsumoto, M.~M.~Nojiri and K.~Tobioka,
  Phys.\ Rev.\ D {\bf 86} (2012) 015005
  [arXiv:1204.2317 [hep-ph]].
  
\bibitem{Girdhar:2012vn} 
  A.~Girdhar and B.~Mukhopadhyaya,
  arXiv:1204.2885 [hep-ph].
  
\bibitem{Vignaroli:2012nf}
  N.~Vignaroli,
  Phys.\ Rev.\ D {\bf 86} (2012) 075017
  [arXiv:1207.0830 [hep-ph]].
  
\bibitem{ATLAS:Tth}
[ATLAS Collaboration], ATLAS-CONF-2013-018.

\bibitem{Berenstein:2012fc} 
  D.~Berenstein, T.~Liu and E.~Perkins,
  arXiv:1211.4288 [hep-ph].
  
\bibitem{Kearney:2013oia} 
  J.~Kearney, J.~Thaler and A.~Pierce,
  arXiv:1304.4233 [hep-ph].

\bibitem{DeSimone:2012fs} 
  A.~De Simone, O.~Matsedonskyi, R.~Rattazzi and A.~Wulzer,
  arXiv:1211.5663 [hep-ph].

\bibitem{Kaplan:1991dc}
  D.~B.~Kaplan,
  Nucl.\ Phys.\ B {\bf 365} (1991) 259.

\bibitem{Agashe:2006at} 
  K.~Agashe, R.~Contino, L.~Da Rold and A.~Pomarol,
  Phys.\ Lett.\ B {\bf 641}, 62 (2006)
  [hep-ph/0605341].

\bibitem{Coleman:1969sm}
  S.~R.~Coleman, J.~Wess and B.~Zumino,
  Phys.\ Rev.\  {\bf 177} (1969) 2239; \\
C.~G.~Callan, Jr., S.~R.~Coleman, J.~Wess and B.~Zumino,
  Phys.\ Rev.\  {\bf 177} (1969) 2247.

\bibitem{Perelstein:2003wd}
  M.~Perelstein, M.~E.~Peskin and A.~Pierce,
  Phys.\ Rev.\ D {\bf 69} (2004) 075002
  [hep-ph/0310039].

\bibitem{Campbell:2009gj}
  J.~M.~Campbell, R.~Frederix, F.~Maltoni and F.~Tramontano,
  JHEP {\bf 0910} (2009) 042
  [arXiv:0907.3933 [hep-ph]].


\bibitem{Aliev:2010zk}
  M.~Aliev, H.~Lacker, U.~Langenfeld, S.~Moch, P.~Uwer and M.~Wiedermann,
  Comput.\ Phys.\ Commun.\  {\bf 182} (2011) 1034
  [arXiv:1007.1327 [hep-ph]].

\bibitem{Martin:2009iq}
  A.~D.~Martin, W.~J.~Stirling, R.~S.~Thorne and G.~Watt,
  Eur.\ Phys.\ J.\ C {\bf 63} (2009) 189
  [arXiv:0901.0002 [hep-ph]].

\bibitem{Atre:2008iu}
  A.~Atre, M.~Carena, T.~Han and J.~Santiago,
  Phys.\ Rev.\ D {\bf 79} (2009) 054018
  [arXiv:0806.3966 [hep-ph]].
  

\bibitem{Alwall:2011uj}
J.~Alwall, M.~Herquet, F.~Maltoni, O.~Mattelaer and T.~Stelzer,
JHEP {\bf 1106}, 128 (2011).

\bibitem{Kidonakis:2011wy}
  N.~Kidonakis,
  Phys.\ Rev.\ D {\bf 83} (2011) 091503
  [arXiv:1103.2792 [hep-ph]].

\bibitem{Kidonakis:2010ux}
  N.~Kidonakis,
  Phys.\ Rev.\ D {\bf 82} (2010) 054018
  [arXiv:1005.4451 [hep-ph]].

\bibitem{Kidonakis:2010tc}
  N.~Kidonakis,
  Phys.\ Rev.\ D {\bf 81} (2010) 054028
  [arXiv:1001.5034 [hep-ph]].

\bibitem{Feynrule}
N.D. Christensen and C. Duhr,
Comput.Phys.Commun. {\bf 180}:1614-1641 (2009).

\bibitem{Sjostrand:2006za}
T.~Sjostrand, S.~Mrenna and P.~Z.~Skands,
JHEP {\bf 0605}, 026 (2006).

\bibitem{Ovyn:2009tx}
  S.~Ovyn, X.~Rouby and V.~Lemaitre,
  arXiv:0903.2225 [hep-ph].

\bibitem{Cacciari:2008gp}
  M.~Cacciari, G.~P.~Salam and G.~Soyez,
  JHEP {\bf 0804} (2008) 063
  [arXiv:0802.1189 [hep-ph]].

\bibitem{Cacciari:2011ma}
  M.~Cacciari, G.~P.~Salam and G.~Soyez,
  Eur.\ Phys.\ J.\ C {\bf 72}, 1896 (2012).

\bibitem{Butterworth:2008iy}
  J.~M.~Butterworth, A.~R.~Davison, M.~Rubin and G.~P.~Salam,
  Phys.\ Rev.\ Lett.\  {\bf 100}, 242001 (2008).

\bibitem{Gallicchio:2010sw} 
  J.~Gallicchio and M.~D.~Schwartz,
  Phys.\ Rev.\ Lett.\  {\bf 105}, 022001 (2010)
  [arXiv:1001.5027 [hep-ph]].
  
\bibitem{Hook:2011cq} 
  A.~Hook, M.~Jankowiak and J.~G.~Wacker,
  JHEP {\bf 1204}, 007 (2012)
  [arXiv:1102.1012 [hep-ph]].
  
\bibitem{BDT}
 B. P. Roe, H. \_J. Yang, J. Zhu, Y. Liu, I. Stancu and G. McGregor, Nul. Instrum. Meth. A 543, 577 (2005).
\bibitem{Hocker}
 A.Hocker et al, PoS ACAT , 040 (2007).

\bibitem{CMS:singletop}
[CMS Collaboration], CMS PAS TOP-12-011.

\bibitem{ATLAS:singletop}
[ATLAS Collaboration], ATLAS-CONF-2012-132.



\end{thebibliography}
\end{document}